%% file: main_arxiv.tex
\PassOptionsToPackage{bookmarksnumbered,unicode}{hyperref}
\RequirePackage{hyperref}

\documentclass[format=acmsmall,review=false,nonacm]{acmart}
\usepackage{booktabs} 
\usepackage[ruled]{algorithm2e} 

\SetAlFnt{\small}
\SetAlCapFnt{\small}
\SetAlCapNameFnt{\small}
\SetAlCapHSkip{0pt}
\IncMargin{-\parindent}
\usepackage{algpseudocode}

\usepackage{csquotes}
\input{common/definitions}
\newtheorem{remark}{Remark}
  
  \newtheorem{imr}{}
\newenvironment{cmr}[1]
  {\renewcommand\theimr{#1}\imr}
  {\endimr}

\usepackage{color}

\setcitestyle{authoryear}

\title{Distribution through Repeated Market with Buying Rights}

 \author{David Sychrovsk\'{y}}
 \affiliation{
 	\institution{Charles University}
 	\city{Prague}
 	\country{Czechia}}
 \email{sychrovsky@kam.mff.cuni.cz}

  \author{Jakub \v{C}ern\'{y}}
 \affiliation{
 	\institution{Columbia University}
 	\country{USA}
 	\city{New York}
    }
 \email{jakub.cerny@columbia.edu}

  \author{Martin Loebl}
 \affiliation{
 	\institution{Charles University}
 	\city{Prague}
 	\country{Czechia}}
 \email{loebl@kam.mff.cuni.cz}

\begin{abstract}
Resource distribution is a fundamental problem in economic and policy design, particularly when demand and supply are not naturally aligned. 
Without regulation, wealthier individuals may monopolize this resource, leaving the needs of others unsatisfied. While centralized distribution can ensure fairer division, it can struggle to manage logistics efficiently, and adapt to changing conditions, often leading to shortages, surpluses, and bureaucratic inefficiencies. Building on previous research on market-based redistribution, we examine a repeated hybrid market that incorporates buying rights. These rights, distributed iteratively by a central authority (for instance, as digital tokens), are intended to enhance fairness in the system — a unit of right is required to acquire a unit of the resource, but the rights themselves can also be traded alongside the resource in the market. We analyze how this regulatory mechanism influences the distribution of the scarce resource in the hybrid market over time. Unlike past works that relied on empirical methods, we explore the exact analytical properties of a system in which traders optimize over multiple rounds. We identify its market equilibrium, which is a natural generalization of the free market equilibrium, and show that it is coalition-proof. To assess the fairness in the system, we use the concept of frustration, which measures the gap between the resources a buyer is entitled to through their buying rights and what they actually obtain through trading. Our main theoretical result shows that using buying rights reduces the frustration by at least half compared to the free market. Empirical evaluations further support our findings, suggesting the system performs well even beyond the theoretically studied assumptions.
\end{abstract}

\begin{document}

\begin{titlepage}

\maketitle


\end{titlepage}

\input{sections/intro2}

\input{sections/02_problemdef}

\input{sections/03_nonmyopic}

\input{sections/04_experiments}

\input{sections/05_conclusion}


\clearpage
\bibliographystyle{ACM-Reference-Format}
\bibliography{references}

\clearpage
\appendix
\input{appendix/00_myopic_case}

\input{appendix/contested_garment}

\input{appendix/02_market_mechanism}

\input{appendix/03_canonical}

\end{document}

%% file: common/definitions.tex
\usepackage{amsfonts}

\makeatletter
\DeclareRobustCommand*\cal{\@fontswitch\relax\mathcal}
\makeatother

\newcommand{\market}{\mathbb{M}}
\newcommand{\crisis}{\mathbb{C}}
\newcommand{\marketFull}{\market^\tau(S, B, G^\tau, M^\tau, D, u, \mu, \phi)}
\newcommand{\crisisFull}{\crisis(S, B, \mathcal{T}, D, u, \mu, \phi, \rho)}

\newcommand{\crisisname}{Repeated Market}
\newcommand{\expectedfrustration}{\ensuremath \mathcal{E}}

%% file: sections/intro2.tex
\section{Introduction}
\label{sec.int}

Policymakers often face the challenge of distributing resources among a population, whether in response to economic planning, social welfare programs, or periods of scarcity. The choice of distribution mechanism can significantly impact fairness, efficiency, and overall societal outcomes. 
Two extreme approaches are centralized distribution by authorities and decentralized distribution through free markets. Markets have the potential to distribute goods flexibly and reliably among many buyers and sellers~\cite{cripps2006efficiency}. However, free markets can lead to significant price increases during scarcity, with powerful individuals acquiring more than their fair share, leaving others with limited or no access to resources. Centralized distribution, on the other hand, while having potential to be fairer, often suffers from economic and time-related inefficiencies~\cite{moroney1997relative}. To address these challenges, policymakers often implement regulatory measures such as price controls, rationing, or subsidies to improve fairness while maintaining efficiency. 

In this paper, we explore to what extent the benefits of both approaches could be combined through the introduction of  \textit{buying rights}. 
We study a repeated hybrid distribution mechanism with money, where participants can trade both the resource and their buying rights. The repetition is a key feature, as the monetary gains from selling buying rights carry over to future iterations, gradually improving the financial standing of lower-income buyers. Each iteration, occurring over discrete time, consists of two stages:
 \begin{enumerate}
    \item \textit{Rights distribution}: Each seller announces the amount of the resource available for sale, while the buyers declare their claims. The central authority then allocates\footnote{Note that new buying rights are issued at each iteration and are valid only during that iteration, expiring at its end.} buying rights to the buyers in an amount equivalent to the total supply of the resource.

    \item \textit{Trading}: Buyers exchange money for the resource and buying rights, subject to the constraint that each unit of the resource requires a corresponding unit of rights. At the end of each iteration, rights expire, and the resource is consumed according to buyers' claims.
\end{enumerate}
To isolate the effects of buying rights, we simplify the market side of the mechanism using an Arrow-Debreu-style framework. Specifically, we assume a centralized\footnote{While the system is centralized, we distinguish this from traditional centralized distribution models that involve significant logistical coordination by a central authority. Here, centralization refers to a platform-based system where rights are distributed by a single authority, but can be transferred and verified digitally -- for example, through token-based ledgers.} system where (i) the resource is a single, divisible commodity, (ii) there are multiple buyers and sellers whose utilities, as well as initial endowments of money, buying rights, and the resource, are public, and (iii) the central authority's rights distribution and sellers-buyers matching mechanisms are also publicly known. 
The iterative market mechanism studied here was originally proposed by Martin Loebl, see \cite{JLS}.
To evaluate the impact of buying rights, we use frustration~\cite{JLS,CFLS}, an adaptation of the Price of Anarchy for systems with buying rights. Frustration is defined as the normalized difference between a buyer’s allocated rights and the actual amount of the commodity they purchase. We argue that, under certain additional assumptions, incorporating buying rights significantly reduces frustration of self-interest traders. Furthermore, we empirically demonstrate that similar improvements occur even in cases where theoretical guarantees are lacking.

The problem of allocating resources in a \enquote{fair} manner has been widely studied over the past decades, with various notions of fairness, such as envy-freeness and maximum Nash welfare, being explored in the literature~\cite{suksompong23characterization}. Many works have investigated the possibility of achieving both fairness and efficiency simultaneously, with a comprehensive survey by \citet{far} providing a detailed overview. Another related area is the theory of claims and taxation problems, particularly fair divisions in bankruptcy scenarios, as surveyed in \cite{T}. A closer connection exists within the literature on redistributive market mechanisms that focus on reducing inequalities. For instance, \citet{DKA} examine a two-sided market for trading homogeneous goods, optimizing the total utility of traders. Our approach differs in that we focus on a fairness measure specific to buying rights and assume that utilities are common knowledge. This line of research has been extended to include settings with heterogeneous qualities of tradable goods, various measures of allocation optimality, and imperfect trader observations \cite{akbarpour2020redistributive}. \citet{condorelli2013market} further explores multiple market and non-market mechanisms for allocating a limited number of identical goods among multiple buyers, arguing that when buyers' willingness to pay aligns with the designer's preferences, market mechanisms are optimal. However, in situations similar to ours -- where the resource is highly valuable to all participants but some lack sufficient funds -- these findings suggest that relying solely on free markets may not be sufficient for achieving the desired distribution. 
As mentioned above, the iterative market mechanism with two-stage structure was introduced in \cite{JLS}, which focuses on myopic traders. 
The idea was further developed in~\cite{CFLS}, where a system akin to the one studied here was investigated, but its equilibria were analyzed only empirically.
The analysis of myopic traders has been developed further in \cite{loebl2025market}, which accompanies this paper.


The system with buying rights that we study also finds parallels in emissions allowances and tradable allowance markets. Historically, regulators directly allocated tradable property rights to firms, which resulted in inefficiencies such as misallocation, regulatory distortions, and barriers to entry. In response, contemporary market designs have shifted to using auctions for the allocation of these rights. As discussed in \cite{dor}, tradable allowance markets play a key role in ensuring the efficiency of carbon markets and preventing market power from being concentrated in the hands of large, dominant agents. The efficiency of multi-round trading auctions for allocating carbon emission rights is further examined in \cite{WCZ}.

\subsection{Organization and contributions}





With these differences in mind, our key contributions are as follows. First, in Section~\ref{sec.def}, we formalize the concept of a repeating trading environment with buying rights and the frustration measure. We explain how combining market and right distribution mechanisms impacts resource redistribution, focusing on the evolution of expected per-round frustration. Second, in Section~\ref{sec:nonmyopic}, we investigate solutions for these trading systems with traders capable of optimizing over a long horizon. We introduce a specific implementation of a single-round Market with such traders and derive an explicit formulation of the equilibrium, which extends the free market’s equilibrium. Theorem~\ref{thm:greedy:equilibrial} provides an algorithm for computing this equilibrium, and we demonstrate it is coalition-proof. Theorem~\ref{thm:nonmyopic:poa} then formulates an upper bound on the expected frustration, showing how the equilibrium shifts toward the desired resource distribution. Finally, in Section~\ref{sec: experiments}, we empirically evaluate the performance of the mechanism, comparing it to free-market scenarios with both constant and time-dependent re-supply of the resource.


%% file: sections/02_problemdef.tex
\section{Problem Statement}

\label{sec.def}





We model the distribution as a repeated market with buying rights. 
We consider one scarce resource, which we call {\it Good}. The other two commodities, representing funds and the right to purchase Good, will be referred to as {\it Money} and {\it Right}.
Each round is an extended market which we call {\it Market}. The traders form two disjoint sets of {\it sellers} and {\it buyers}. Sellers engage in selling Goods, while buyers can participate in selling and buying Rights, and purchasing Goods. 
Each single-round Market 
starts with the sellers declaring the amount of Good for sale. 
The buyers declare their {\it Claim} for Good, which remains fixed across the repetitions. 
A central {\it rights distribution mechanism} then assigns buyers with appropriate amount of Right. At the beginning of each Market, each seller and buyer also receive an amount of Good and Money, respectively. 
Our theoretical results require these amounts to be constant across the entire sequence of repeated Markets. 
The traders then trade the Good, Right and Money. 
An important restriction is that {\em at the end of each Market, each buyer has at least as much Right as Good.} This restriction forms the core of our approach. 

\begin{remark}
\label{rem: money next market}
We further require that Money obtained by selling Right cannot be used in the current Market for buying Good. This rule is designed to allow the wealthier buyers to obtain the Good first. All our results hold, typically with simpler proofs\footnote{When the Money obtained for selling Right can be used in the current Market, the whole sequence of \crisisname{} decomposes into isolated instances of Markets. Each Market can be solved analytically, see Appendix~\ref{app: myopic case}.}, even without this assumption. Another ethical requirement is that the Money obtained by selling Right stay in the system and is used to purchase Good in the future.
\end{remark}

We refer to the single-round trading that happens in a Market after the Right is distributed as {\em Trading}. Each Market hence consists of the buying rights distribution phase and Trading. At the end of each Market, (1) buyers consume all the obtained Right, (2) buyers consume all the obtained Good up to their declared Claim and keep the surplus Good and obtained Money to the next Market, and (3) sellers consume the obtained Money.
The Markets are thus interdependent.


\subsection{Single-round Market}
\label{sec: trading env}

Formally, in a timestep $\tau\in\{1,\dots \mathcal{T}\}$, a Market is a tuple $\mathbb{M}^\tau= \marketFull$, where $S$ is a set of sellers and $B$ is a set of buyers. 
We denote $T = S \cup B$ the set of all traders and assume $S \cap B = \emptyset$. 
The $G^\tau = \{G_t^\tau|t\in T\}$ and $M^\tau = \{M_t^\tau|t\in B\}$ denote the sets of Good and Money each trader has at the beginning of a Market $\tau$, respectively. 
We also denote subsets of a set with a subscript, for example $G^\tau_A = \{G^\tau_a|a\in A\}$ for a set $A\subset T$. The set of Claim $D = \{D_b|b\in B\}$ gives the amount of Good each buyer hopes to acquire in the Market. 
The functions $u$ and $\mu$ denote the \emph{utility} of traders and \emph{market mechanism}, respectively, and will be defined formally later in Section~\ref{sec:nonmyopic}.  
Finally, $\phi$ is the (rights) distribution mechanism. 
We let $V^\tau \leq \sum_{s\in S} G_s^\tau$ be the offered volume of Good in $\mathbb{M}^\tau$. 

\begin{definition}
\label{def.fairness}
A right distribution mechanism is a function $\phi: \mathbb{R}^{+, |B| + 1}_0 \to \mathbb{R}^{+, |B|}_0$ which, given the total offered volume of Good $V$ 
 and Claim $D_B$ of buyers, assigns Right to the buyers s.t. $\forall V, V' \in \mathbb{R}_0^+, D_B\in\mathbb{R}^{+,|B|}_0,\forall b\in B$ it holds 
\begin{enumerate}
    \item $\sum_{b\in B}\phi_b(V, D) = V$,
    \item $D_b\ge D_b'\Rightarrow\phi_b(V, D) \ge \phi_b(V, D')$,
    \item $V \ge V' \Rightarrow \phi_b(V, D) \ge \phi_b(V',D)$.
\end{enumerate}
\end{definition}

Note that we left the definition as non-restrictive\footnote{One restriction we make is that the distribution mechanism is myopic and non-strategic. While a central planner could, in principle, allocate rights in each round in a more forward-looking manner---anticipating how rights will be traded and optimizing outcomes over a longer horizon---we make a conscious decision not to do so. This reflects practical constraints: the desire for transparency and ethical clarity in allocation, as well as the difficulty of defining or justifying a long-term planning horizon.}
 as possible. 
Depending on the application, one might require additional properties to capture the notion of ``fairness''.
For example, $D_b=0\Rightarrow\phi_b(V, D) = 0$, or anonymity, defined in our context as 
\begin{equation*}
    \phi_b(V, \alpha(D)) = \phi_{\alpha^{-1}(b)}(V, D),
\end{equation*}
for all permutations $\alpha$ of $|B|$ elements.
%
Our results do not depend on a particular choice of a distribution mechanism. 
We give two examples, which we will use also later in the empirical evaluation in Section~\ref{sec: experiments}: 
the {\it proportional fairness mechanism}, which allocates Right to $b\in B$ proportionally to their Claim, i.e., 
\begin{equation}
    \label{eq: proportional distribution}
    \phi_b\left(V, D\right) =
\frac{D_b V}{\sum_{b\in B} D_b},
\end{equation} 
and the {\it contested garment distribution}, designed to fairly resolve conflicting claims~\cite{AM}. Formal description of the contested garment distribution is not required for the exposition; for more details see Appendix~\ref{app: contested garment}.


Assigning the $\phi_b(V^\tau, D)$ amount of Right to each buyer~$b$ by the distribution mechanism~$\phi$ constitutes the first of the two steps of a Market number~$\tau$, $\mathbb{M}^\tau$. In the second step, traders trade assigned Good, Right and Money in the Trading. Recall that Trading is a market with two restrictions: (1) the final basket of each buyer has at least as much Right as the amount of Good and (2) Money obtained for selling Right cannot be used to buy Good in the current Trading. Also, as remarked earlier, the subsequent Markets are not independent; a feasible solution of $\market^\tau$ influences initial endowments of Money in $\market^{\tau+1}$.

 \subsection{Frustration per-buyer and in expectation}
 \label{sec.fair}

The concept of the Right can be understood as the socially determined entitlement of a buyer to a specific amount of Good. Put simply, if the distributional crisis were entirely managed by a central authority, the buyer would receive precisely that amount of Good. The concept of \enquote{frustration} captures the contrast between this ideal centralized solution and the actual quantity a buyer is able to obtain.

Formally, the frustration of a buyer is the normalized difference between the Right he would be assigned and the amount of Good he has in a Market if that is at least zero, and zero otherwise. 
We note that this definition allows measuring frustration also in the free market, where no Right is assigned. 
However, a common notion of desirable allocation may still exist. 
Minimization of frustration captures the objective of moving towards the desirable distribution.
 
\begin{definition}
\label{def.fru}
Let $V^\tau$ be the amought of offered Good in a Market $\tau$, $D$ be the Claim, and $G_b^\tau$ be the amount of Good $b\in B$ has after $\market^\tau$. 
Then the {\em frustration} of buyer $b\in B$ is 
\begin{equation*}
    f_b^\tau = 
    \max\left\{0, \frac{\phi_b(V^\tau, D)- G_b^\tau}{\phi_b(V^\tau, D)}\right\},
\end{equation*}
if $\phi_b(V^\tau, D)>0$, and zero otherwise.
\end{definition}

When the Good is traded, the final allocation may differ from the centralized distribution. This disparity serves as a measure of the inefficiency inherent in trading when it comes to allocating the Good in a manner that aligns with the central authority's preferences. This concept bears resemblance to the Price of Fairness or the Price of Anarchy, which quantifies the cost incurred by the system due to the autonomous behavior of the involved actors~\cite{koutsoupias1999worst}. We define the \emph{expected frustration} in our system as a scaled sum of frustrations of the buyers
\begin{equation*}
    \expectedfrustration_f^\tau = \frac{1}{\tau |B|}\sum_{i=1}^\tau\sum_{b\in B}f_b^i.
\end{equation*}
Note that $\expectedfrustration_f^\tau \in [0, 1]$ (since $f_b^\tau \in [0, 1]$), and $\expectedfrustration_f^\tau = 0 \Leftrightarrow f_b^i = 0 \ \forall b\in B, \forall i\in \{1,\dots \tau\}$. In the latter case, trading Good in the Market has the same social impact as distributing it centrally.

%% file: sections/03_nonmyopic.tex
\section{Repeated Market with Strategic Long-Term Traders}
\label{sec:nonmyopic}
Now we move to the investigation of solutions of the \crisisname{} where traders optimize over the entire horizon. 
We begin our analysis by introducing a specific implementation of a single-round Market with such traders.
We derive an explicit formulation of the solution in the form of the interaction's equilibrium, which turns out to be a simple extension to the free market's equilibrium. 
Furthermore, we examine its robustness with respect to coalitions, and formulate an upper-bound on the arising expected frustration. 

\label{ssec: game theoretic formulation}




Formally, the Market $\market^\tau$ is given by the set of all possible \emph{strategies} of the traders, the \emph{market mechanism}, and the \emph{utilities} of the traders. 
Let us address each of these components separately, and then focus on their interplay. 
The sellers are motivated by profit, so their utility is the amount of Money they acquire in a Market. We modify this simple model by subtracting the amount of Good they are left with after the Market. This represents the cost of storing, as well as the damaged reputation by not selling the scarce Good. Together,
\begin{equation*} \label{eq: seller utility}
    u^\tau_s(G_s^\tau, M_s^\tau) =
    M_s^\tau - c\ G_s^\tau, 
\end{equation*} 
where $c > 0$ is a suitable constant and $G_s^\tau, M_s^\tau$ are the initial values modified by the market mechanism $\mu$, see Definition~\ref{def: informal market mechanism}. A buyer, on the other hand, wishes to ensure a steady supply of Good during the entire sequence of repeated Markets. 
We assume each buyer makes use only of the Good they consume at the end of the Market via 
\begin{equation*} \label{eq: buyer utility}
    u^\tau_b(G_b^\tau, M_b^\tau) =
    \min\left\{D_b, G_b^\tau\right\}.
\end{equation*}
The buyers can only receive utility for Money indirectly, through the Good they acquire by participating in the repeated Market.

We denote by $\Pi$ the set of (pure) strategies of the traders. We assume each seller has information about the amount of Good every seller has, as well as Money and Claim\footnote{Since the distribution mechanism is assumed to be public knowledge, the traders also know (for some offered volume of Good) the amount of Right each buyer will be assigned.} each buyer has. 
On the other hand, a buyer has access only to the amount of Money and Good they already own, as well as their Claim. 
We consider this assumption sufficiently realistic as the sellers may invest in some market research, while the such information remain rather limited to the buyers. The strategy of a seller is thus a function $\pi_s: \mathbb{R}^{+, 2|B| + |S|}_0 \to \mathbb{R}^{+, 2}_0$, denoted as
$\pi_s(G_B^\tau, M_B^\tau, G_S^\tau) = (v_s^\tau, p_s^\tau)$,
where $v_s^\tau \le G_s^\tau$ is the offered volume of Good at price $p_s^\tau$ by seller $s$ in Market $\tau$. Based on the total offered volume of Good $V^\tau = \sum_{s\in S} v_s^\tau$ and Claim of buyers, the distribution mechanism $\phi$ allocates the Right. 
When offering Right for sale, a buyer is given the offers of sellers and the amount of Good, Money and Right they have. Their strategy for this task is hence a function $\hat{\pi}_b: \mathbb{R}^{+, 2|S| + 3}_0 \to \mathbb{R}^{+, 2}_0$, denoted as
$\hat{\pi}_b(v_S^\tau, p_S^\tau, G_b^\tau, M_b^\tau, R_b^\tau) = (w_b^\tau, q_b^\tau),$
where $w_b^\tau \le R_b^\tau$ is the offered volume of Right at price $q_b^\tau$. 
When declaring acceptable price and volume for both Good and Right, a buyer is also given the offers of the other buyers. 
Summarizing, the (complete) strategy of a buyer $b$ is a function $\pi_b: \mathbb{R}^{+, 2|S| + 3 + 2(|B| - 1)}_0~\to~\mathbb{R}^{+, 6}_0$, written as $\pi_b(v_S^\tau, p_S^\tau, G_b^\tau, M_b^\tau, R_b^\tau, w_{-b}^\tau, q_{-b}^\tau) = (w_b^\tau, q_b^\tau, \overline{v}_b^\tau, \overline{p}_b^\tau, \overline{w}_b^\tau, \overline{q}_b^\tau)$,
where $\cdot_{-b}^\tau = \{\cdot_{b'}^\tau|b'\in B\setminus \{b\}\}$ and $\overline{\cdot}_b^\tau$ denote acceptable amounts of the corresponding quantities for buyer $b$. We will denote $\pi = \pi_S \times \pi_B$ the strategy profile of all traders. After all traders declare their bids, the Trading begins, which is done by the following market mechanism.

\begin{definition}[market mechanism\footnote{See Appendix~\ref{app: market mechanism} and Algorithm~\ref{alg: proportional market mechanism} for more details.}]
\label{def: informal market mechanism}
The market mechanism is a function $\mu: \Pi\times\mathbb{R}^{+, 2|T|+|B|}_0 \to \mathbb{R}^{+, 2|T|}_0$, written in its entirety as $\mu(v_S^\tau, p_S^\tau, w_B^\tau, q_B^\tau, \overline{v}_B^\tau, \overline{p}_B^\tau, \overline{w}_B^\tau, \overline{q}_B^\tau, G_T^\tau, M_T^\tau, R_B^\tau) = (G_T^\tau, M_T^\tau)$.
The market mechanism is implemented in two stages. In the first stage, the buyers use the Right from their initial endowment to buy Good up to their limit $\overline{v}_b^\tau$, or until they exhaust their Right or Money. 
In the second stage, the buyers buy Good and Right with Money in equal volume, until they buy their desired volume of either, or are left with no Money. 
In both stages, items offered at a lower price are traded first. 
When more traders offer Good or Right at the same price, they sell them at an equal rate. 
\end{definition}





\medskip

\begin{figure*}[t]
\centering
\includegraphics[width=0.95\linewidth]{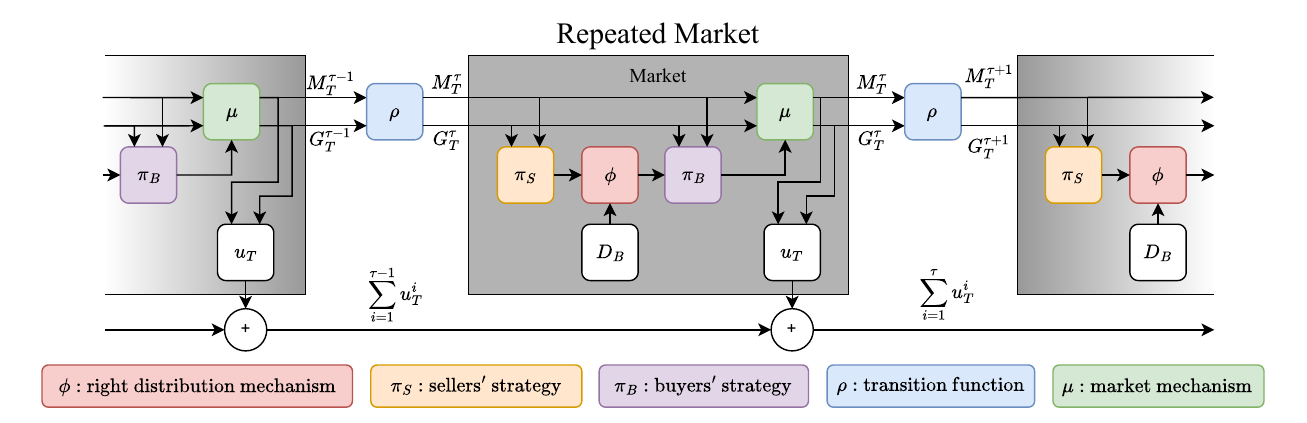}
\caption{An illustration of a \crisisname~consisting of a series of Markets indexed by $\tau$. The distribution function $\phi$ is followed by the Trading phase, which is done through the market mechanism $\mu$. The Money and Good obtained by traders are transferred to the next Market via the transition function $\rho$. The \enquote{+} operation is used to aggregate the utility $u_T^\tau$ of traders $T$.
}
\label{fig: crisis}
\end{figure*}

We are interested in the performance of the system over the period of multiple rounds, each realized by a Market. 
We refer to the sequential game consisting of $\mathcal{T}\in\mathbb{N}$ Markets as the \textit{\crisisname}, see Figure~\ref{fig: crisis} for an illustration. 
Formally, the \crisisname~is a tuple $\crisis(\mathcal{T}) = \crisisFull$, where $S,B,D,\phi$ and $\mu$ have the same meaning as before, $\mathcal{T}$ is the horizon, and $\rho$ is a {\it transition} function which formalizes the consumption of resources\footnote{Formally, we also need to remove Money of sellers to prevent them from obtaining utility for it multiple times.} after each Market, and give traders Good and Money for the next Market.

\begin{definition}
\label{def: transition function}
The transition function $\rho: \mathbb{R}_0^{+, 2}\to \mathbb{R}_0^{+, 2}$ for a seller $s\in S$ and a buyer $b\in B$ is given by
\begin{align*}
    \rho (G_s^\tau, M_s^\tau) &= 
    (g_s + G_s^\tau, 0),\\
    \rho (G_b^\tau, M_b^\tau) &= 
    (\max\{0, G_b^\tau - D_b\}, m_b + M_b^\tau),
\end{align*}
where $g_s, m_b \ge 0$ are the amount of Good and Money sellers and buyers regularly gain after each Market respectively. 
\end{definition}

Our theoretical results rely on the assumption that the quantities $g_s$ and $m_b$ are \emph{constant} during the \crisisname\footnote{This assumption is not only crucial for our analysis but also, we believe, practical in real-world scenarios where rationing through buying rights typically occurs, such as during wartime resource allocation, environmental cap-and-trade systems, or emergency medical supply distribution.}. 
We relax this assumption in some of our experiments, see Section~\ref{ssec: time dependent supply}.
Without a loss of generality, we set $\sum_{s\in S} g_s = 1$ and $\sum_{b\in B} m_b = 1$. 
We further assume the traders aggregate their utilities over the whole \crisisname, up to the horizon $\mathcal{T}$, so that their resulting total utilities are
\begin{equation*}
    u_t \coloneq \sum_{\tau=1}^\mathcal{T} u_t^\tau. 
\end{equation*}
As a solution of this interaction, we employ the standard notion of equilibrial strategies, such that no deviation of a trader may increase their utility. 

\begin{definition}
\label{def: sequence equilibrium}
For a \crisisname~$\crisis(\mathcal{T})$, a strategy $\pi\in\Pi$ is an equilibrium, if for any strategy profile $\overline{\pi}_t\in\Pi$ and any $t\in T$
\begin{equation*}
    u_t(\pi) \ge 
    u_t(\overline{\pi}_t, \pi_{-t}), \hspace{5ex} \forall t\in T,
\end{equation*}
where $u_t(\pi) = \sum_{\tau=1}^\mathcal{T} u_t^\tau(\pi)$, and $u_t^\tau(\pi)$ is the utility received by trader $t$ in Market $\tau$ under strategy profile $\pi$.
\end{definition}

Note that the \crisisname~can be defined with an infinite horizon $\mathcal{T}\to\infty$ as well.
In such a case, since $u_t^\tau$ is bounded, we could use, e.g., exponential discounting to make the total utility finite and the equilibrium well defined~\cite{sutton2018reinforcement}.



\subsection{Greedy strategy}
\label{ssec: gredy strategy}

At first glance, it may be unclear how the inclusion of Right in trading affects the strategies of rational traders. 
We will demonstrate that a particular equilibrium of the \crisisname{} is a natural extension of the free market equilibrium, with no greater expected frustration. 
Informally, this proposed strategy is that the sellers post the highest price that the buyers can afford to pay, taking into account the cost of Right. 
As a result, buyers will purchase all available Good, which also means some might purchase Right. 
Those who sell Right will do so for the same price as for the Good. 
Buyers will not accept prices higher than this for either Right or the Good. 
Finally, sellers will offer exactly the amount of Good $g_s$ that they receive at the beginning of a Market, see Remark~\ref{remark: pseudo-perisable good}. 
We refer to such strategies as \textit{Greedy}, and they will be our focus from now on.

\begin{definition}
\label{def: greedy strategy}
The Greedy strategies of the sellers are
\begin{align}
    \label{eq: seller policy}
    \pi_s(G_B^\tau, M_B^\tau, G_S^\tau) &= 
    \left(g_s, p^\tau\right),
\end{align}
where $p^\tau$ is the solution of
\begin{equation}
    \label{eq: def implicit price}
    \sum_{b\in B}M_b^\tau - \max\{0, p^\tau R^\tau_b - M^\tau_b\}= 
    p^\tau\sum_{b\in B}  R^\tau_b.
\end{equation}
For buyers, the Greedy strategies are
\begin{equation}  
\label{eq: buyer policy}
\pi_b(v_S^\tau, p_S^\tau, G_b^\tau, M_b^\tau, R_b^\tau, w_{-b}^\tau, q_{-b}^\tau) = 
\left(\psi_b^\tau, P^\tau, R_b^\tau+\xi_b^\tau, P^\tau, \xi_b^\tau, P^\tau\right),
\end{equation}
where $P^\tau = \sum_{s\in S}p^\tau_s / |S|$ is the average selling price of Good, $\psi_b^\tau \coloneq \max\{0, R_b^\tau - M_b^\tau / P^\tau\}$, and $\xi_b^\tau \coloneq \max\{0, M_b^\tau / P^\tau - R_b^\tau\}$.
\end{definition}

\begin{remark}[Correctness of Definition \ref{def: greedy strategy}]
It may not be immediately clear how a seller can derive $p^\tau$ from Eq.~(\ref{eq: def implicit price}), as it requires knowledge of $R_B^\tau$. 
But to assign the Right to buyers, the distribution mechanism $\phi$ needs to know the offered volume of Good.
However, if all sellers adopt the Greedy strategy, the offered volume is $V = \sum_{s\in S}g_s$, and, since the distribution mechanism $\phi$ is publicly known, each buyer can compute $R_B^\tau$. 
In contrast, a buyer $b$ does not know $M^\tau_{-b}$, and must therefore estimate the price of the Right by averaging the selling price of the Good.
\end{remark}

\begin{remark}[To Intentional Storing]
\label{remark: pseudo-perisable good}
The Greedy strategy for a seller $s$ is to offer the amount of Good they receive at the beginning of each Market, $g_s$, as if the Good was perishable. At first, this seems to force sellers to store Good when they have surplus from the previous Market. 
But that is not the case as, if all traders follow the Greedy strategy, all Good is sold. 
However, it prevents certain deviations of buyers from Greedy to be beneficial. 
\end{remark}

We justify the exact form of Eq. (\ref{eq: def implicit price}) in the next section. 
Eq.~\eqref{eq: buyer policy} states that, under the Greedy strategy, a buyer $b$ will offer Right at the price of Good offered by sellers, and will not accept prices higher than this. 
He will offer $R_b^\tau - M_b^\tau / P^\tau$ units of Right for sale if that is at least zero, as that is exactly the amount he cannot use after buying Good with his Money. 
Alternatively, if he has Money after exhausting all Right, he will buy $M_b^\tau / P^\tau - R_b^\tau$ units of Right.


\subsection{Greedy strategies form an equilibrium of the \crisisname}
\label{sec: equilibrium}




We will now demonstrate how the Market operates when all traders follow their Greedy strategies. 
All sellers and buyers post the same selling prices, $p^\tau$ and $q^\tau$, respectively, for the Good and Right. 
The amount of Good that all sellers can sell in the first stage is $\sum_{b\in B} \text{min}\{R^\tau_b, M_b^\tau / p^\tau\}$. 
We will call a buyer {\em rich} in a Market $\market^\tau$ if $p^\tau R^\tau_b < M^\tau_b$, and {\em poor} otherwise. 
In the second stage, poor buyers will offer all the Right that they cannot use in the first stage, as they know the prices set by sellers before offering Right.
The amount of offered Good is $\sum_{b\in B} R^\tau_b$, which means the amount of Good sold in the second stage is
\begin{equation*}
    \sum_{b\in B} R^\tau_b - \min\{R^\tau_b, M_b^\tau / p^\tau\} = 
    \sum_{b\in B} \max\{0, R^\tau_b - M^\tau_b / p^\tau\}.
\end{equation*}

The amount of Money sellers get from both stages is 
$ \Delta M_S^\tau = \sum_{b\in B} p^\tau R^\tau_b$,
since all Good is sold. 
Note that this is less than or equal to the amount of Money the buyers posses in total because the Money used for selling Right may only be used in the following Markets. That Money is the amount of Right sold in the second stage times the price
\begin{equation}
\label{eq: useless money}
    \Delta M_B^\tau = \sum_{b\in B} q^\tau\max\{0, R^\tau_b - M^\tau_b / p^\tau\} \ge 0,
\end{equation}
where the equality holds only when $ p^\tau R^\tau_b = M^\tau_b, \forall b\in B$, or when all buyer have enough Right to buy all Good they desire. 
The inequality of the buyers thus dictates how much of the Money can be used to buy Good. 
We will refer to $\Delta M_S$ as {\it useful Money} in a Market and to $\Delta M_B$ as {\it useless Money}. The total Money in the system is the sum of both, i.e.,
\begin{equation*}
\sum_{b\in B} M^\tau_b = 
\sum_{b\in B} p^\tau R^\tau_b + q^\tau \max\{0, R^\tau_b - M^\tau_b / p^\tau\}.
\end{equation*}

We can also derive this equation by taking into account that the cost of the Good is $p^\tau$ in the initial stage and $p^\tau + q^\tau$ in the subsequent stage. This implies that 
\begin{equation}
    \label{eq: implicit price}
    \sum_{b\in B}M_b^\tau - \frac{q^\tau}{p^\tau}\max\{0, p^\tau R^\tau_b - M^\tau_b\}= 
    p^\tau\sum_{b\in B}  R^\tau_b,
\end{equation}
which offers a nice intuition: the Money each buyer has is effectively decreased by the (scaled) frustration which $p^\tau$ would induce. This is because the rich buyers will buy the amount of Right exactly equal to the frustration of the poor, hence the scaling by the relative price. If the traders follow Greedy, then $q^\tau=p^\tau$ and Eq. (\ref{eq: implicit price}) reduces to Eq. (\ref{eq: def implicit price}).

\begin{remark}[To the Money in the Next Market]
In the following Market, if traders follow Greedy, the useless Money is allocated to buyers who were poor in the previous Market. Specifically, in the following Market a buyer $b$ will have
\begin{equation} \label{eq: next money}
    M^{\tau + 1}_b =
    m_b + q^\tau R_b^\tau f_b^\tau
    =
    m_b + 
    \max\{0, p^\tau R^\tau_b - M^\tau_b\}.
\end{equation}
In other words, $b$ earns a portion of the Money proportional to his frustration $f_b^\tau$ (see Definition \ref{def.fru}) in this Market.  
\end{remark}



We are now ready to state the main result of this section.

\begin{theorem}\label{thm:greedy:equilibrial}
The Greedy strategies given in Eq. \eqref{eq: seller policy} and \eqref{eq: buyer policy} form an equilibrium of a \crisisname~of any length. 
Furthermore, the equilibrium is coalition-proof and can be computed in polynomial time in the number of buyers and sellers.
\end{theorem}

\input{appendix/01_proof_equilibrium}

When traders follow the Greedy strategy, all Goods are sold, and the amount of Goods offered remains the same in every Market. 
Therefore, the amount of Right allocated to any buyer doesn't change either.
It is also worth noting that, under these assumptions, $\sum_{b\in B} R_b^\tau = 1$. 
We present a full analytic solution of our hybrid system for a particular right distribution mechanism in Appendix~\ref{app: canonical solutions}.

\subsection{Upper-bounding the frustration}

One may observe certain similarities between Eq. (\ref{eq: implicit price}) and (\ref{eq: next money}). 
The frustration in $\market^\tau$ decreases useful Money a buyer has, and increases the amount they will have in the following Market. 
We will now show that the price oscillates around a fixed value and tends to it over time. 
Moreover, that fixed value is the free market clearing price.
\begin{proposition}\label{thm:nonmyo:main}
Let all traders follow the Greedy strategy. Then the mapping of the current price to the next is a non-expansive mapping on $\mathbb{R}$ with the $L_1$ norm. Furthermore, the price of Good and Right approaches one as $\tau\to\infty$.
\end{proposition}


\input{appendix/01_proof_nonexpansive}

This suggests that the \crisisname~ultimately reaches a stable state. 
Additionally, the price is the same as it would be in a free market $\frac{\sum_{b\in B}m_b}{\sum_{s\in S}g_s} = 1$. 
This means the distribution mechanism primarily impacts the income of buyers.
In this scenario, since $\lim_{\tau \to \infty} p^\tau~=~1$, the amount of Money and Goods entering and leaving the system in each Market are equal.
Importantly, the sellers don't have any incentive to seek alternative ways of selling good outside our hybrid market. 
This is because under the Greedy strategy, they obtain all the Money of the buyers, just like in the free market.
This stands in stark contrast to the centralized distribution, which typically forces the sellers to use below market-clearing prices.

The amount of Money~\eqref{eq: next money} buyers start a Market with hence asymptotically stabilizes to
\begin{equation}\label{eq: money in limit}
M_b = m_b + \max\{0, R_b - M_b\}.  
\end{equation}
This leads to the expected frustration being twice as high in the \crisisname~with the free market than in our system.

\begin{theorem}\label{thm:nonmyopic:poa}
Consider a \crisisname~where traders follow the Greedy strategy. Then the expected frustration in the Market with the right distribution mechanism is at most $1/2$ of the free market's expected frustration as the Market number $\tau \to \infty$.
\end{theorem}


\input{appendix/01_proof_limit_price}

%% file: appendix/01_proof_equilibrium.tex
\begin{proof}
We prove the theorem as a sequence of lemmas. We begin with two useful lemmas, showing how the price changes with Money, and throughout the \crisisname{}.

\begin{lemma}
\label{lemma: increasing price}
Let the traders follow the Greedy strategy. Then $p^{\tau}$, given as a solution of Eq. (\ref{eq: def implicit price}) is an increasing function of $M^\tau_b$  $\forall b\in B, \tau\in\{1, \dots {\cal T}\}$.
\end{lemma}
\begin{proof}
Taking the derivative of both sides yields
\begin{align*}
    \frac{{\rm d} p^\tau}{{\rm d} M_{b'}^\tau} 
    &=
    \frac{{\rm d} }{{\rm d} M_{b'}^\tau}\sum_{b\in B}M^\tau_b - \max\{0, p^\tau R_b - M^\tau_b\} \\
    &=1-\sum_{b\in B}\text{sign}(p^\tau R_b - M^\tau_b)\left(\frac{{\rm d} p^\tau}{{\rm d} M_{b'}^\tau}R_b - 1\right) \\
    &=1+\tilde{N}-\frac{{\rm d} p^\tau}{{\rm d} M_{b'}^\tau}\sum_{b\in B}\text{sign}(p^\tau R_b - M^\tau_b)R_b,\\
    \hspace{1ex}&\Rightarrow 
    \frac{{\rm d} p^\tau}{{\rm d} M_{b'}^\tau} = 
    \frac{1+\tilde{N}}{1+\tilde{R}} \ge 1,
\end{align*}
where $\tilde{N}>0$ is the number of poor buyers and $0<\tilde{R}<\tilde{N}$ is the sum of their Right.
\end{proof}

\begin{lemma}
\label{lemma: next price}
Let the traders follow the Greedy strategy. Then $p^{\tau-1} < 1 \Rightarrow p^\tau > p^{\tau-1}$, resp. $p^{\tau-1} > 1 \Rightarrow p^\tau < p^{\tau-1}$.
\end{lemma}
\begin{proof}
Substituting Eq. (\ref{eq: next money}) into (\ref{eq: implicit price}) gives
\begin{align}
\nonumber
    \sum_{b\in B}m_b + \max\{0, p^{\tau-1} R_b - M^{\tau-1}_b\} 
    - \max\{0, p^\tau R_b - M^\tau_b\} &= 
    p^\tau, \\
    1 + \sum_{b\in B}\max\{0, p^{\tau-1} R_b - M^{\tau-1}_b\}  
    \label{eq: price oscillation}
    - \max\{0, p^\tau R_b - M^\tau_b\} &= 
    p^\tau,
\end{align}

If $p^{\tau - 1} < 1$, then the useful Money also is
\begin{equation*}
\sum_{b\in B} M^{\tau-1}_b - \max\{0, p^{\tau-1} R_b - M^{\tau-1}_b\} < 1,
\end{equation*}
or in other words
\begin{equation*}
\sum_{b\in B} M^{\tau-1}_b
< 
1 + \sum_{b\in B} \max\{0, p^{\tau-1} R_b - M^{\tau-1}_b\}.
\end{equation*}
Going back to Eq. (\ref{eq: price oscillation}), we see that
\begin{align*}
    1 + \sum_{b\in B}\max\{0, p^{\tau-1} R_b - M^{\tau-1}_b&\} 
    = 
    p^\tau + \sum_{b\in B}\max\{0, p^\tau R_b - M^\tau_b\},\\
    \sum_{b\in B} M^{\tau-1}_b <
    p^\tau + \sum_{b\in B}\max\{0, p^\tau &R_b - M^\tau_b\} 
    = \sum_{b\in B} M^{\tau}_b.
\end{align*}
Using Lemma \ref{lemma: increasing price} we get $p^\tau > p^{\tau - 1}$. Similarly, we can show $p^{\tau-1} > 1 \Rightarrow p^\tau < p^{\tau - 1}$.
\end{proof}

To demonstrate that the Greedy strategy is an equilibrium, we show that no deviation from it is beneficial for any trader. We consider changes in the offered volume of Good and Right first.

\begin{lemma}
\label{lm:nonmyo:utildec}
Let all traders follow Greedy except for a seller who offers less Good in a Market and sells it in the following Market. Then his utility decreases as a consequence.
\end{lemma}
\begin{proof}
Note that the scenario in which the seller deviates is denoted with a hat. Since the distribution mechanism is non-decreasing in the total offered volume, we get $\hat{R}_b^{\tau+1} \ge R_b^{\tau+1}$ for all buyers. For contradiction we also assume the deviation is beneficial for the seller, i.e. $\hat{p}^{\tau+1} \ge p^{\tau+1}$. Since at least one buyer didn't get the Money for selling Right in $\tau$, we have
\begin{equation}
\label{eq: frustration ineq}
    \sum_{b\in B}\max\{0, \hat{p}^{\tau+1} \hat{R}^{\tau+1}_b - \hat{M}^{\tau+1}_b\}\\
    >
    \sum_{b\in B}\max\{0, p^{\tau+1} R^{\tau+1}_b - M^{\tau+1}_b\}.
\end{equation}
Using Eq. (\ref{eq: def implicit price}) in both scenarios gives
\begin{align}
    \label{eq: price no hat}
    \sum_{b\in B}M_b^{\tau+1} - \max\{0, p^{\tau+1} &R^{\tau+1}_b - M^{\tau+1}_b\}= 
    p^{\tau+1},\\
    \label{eq: price hat}
    p^\tau {\cal V} + \sum_{b\in B}M_b^{\tau+1} - \max\{0, \hat{p}^{\tau+1} &\hat{R}^{\tau+1}_b - \hat{M}^{\tau+1}_b\}
    = 
    \hat{p}^{\tau+1}(1+{\cal V}),
\end{align}
or in combination with Eq. (\ref{eq: frustration ineq})
\begin{equation*}
    p^\tau {\cal V} + p^{\tau + 1} >
    \hat{p}^{\tau+1}(1+{\cal V}).
\end{equation*}
The price if $s$ deviates is thus upper bounded by the weighted average of prices under Greedy. We can consider two cases.
\begin{enumerate}
    \item $p^\tau < 1$: Then by Lemma \ref{lemma: next price}, $p^{\tau + 1} \ge p^\tau \Rightarrow p^{\tau + 1} > \hat{p}^{\tau+1}$, which is a contradiction.
    \item $p^\tau > 1$: Again by Lemma \ref{lemma: next price}, $p^{\tau + 1} \le p^\tau \Rightarrow p^\tau > \hat{p}^{\tau+1}$. This means the seller could have sold Good in $\tau$ and increase his payoff.
\end{enumerate}
\end{proof}

\begin{lemma}
\label{lemma: buyer deviation}
Let all traders follow Greedy except for a rich, resp. a poor buyer who buys, resp. sells less Right in a Market. Then his utility decreases as a consequence. 
\end{lemma}
\begin{proof}
If a buyer deviates in this way, it leaves sellers with more Good for the next Market, and the buyers with more Money. But when following Greedy, the sellers will offer the same volume of Good in the next, leading to the same distribution of Right as in the last Market. And, since the buyers now have more Money than by Lemma \ref{lemma: increasing price}, the price increases. Moreover, if $b$ was poor in $\tau$, then he would receive less Money from selling Right, limiting the amount of Good he can buy in the next Market
\begin{equation*}
    \hat{M}^{\tau+1}_b = m_b + \max\{0, p^{\tau} (R_b-{\cal V}) - M^{\tau}_b\} < M^{\tau+1}_b.
\end{equation*}
\end{proof}

Deviating from Greedy by changing the price has a similar effect.

\begin{lemma}
\label{thm: equal prices}
Let all traders follow the Greedy strategy. Then no trader can increase his utility by changing the selling price of Good, resp. Right in a single Market.
\end{lemma}
\begin{proof}
Let us start with $s\in S$ changing $p^\tau_s$. There are two cases
\begin{enumerate}
    \item $p^\tau_s < p^\tau$: In this situation, $s$ will get less Money in $\tau$, which will stay in the system. However, in the following Market, the Money will be split proportionally to $g_s$, decreasing the utility of $s$.
    \item $p^\tau_s > p^\tau$:
    The acceptable price of Good is the average of the selling prices. This means $s$ will not sell anything, decreasing his utility as was shown in Lemma \ref{lm:nonmyo:utildec}.
\end{enumerate}
Similarly for $b\in B$
\begin{enumerate}
    \item $q^\tau_b < q^\tau$: When selling at a lower price, $b$ will get less Money in the next Market, decreasing his utility. Furthermore, some rich buyers will be left with more Money, increasing the price in the following Market.
    \item $q^\tau_b > q^\tau$:
    The acceptable price of Right is $p^\tau=q^\tau$, so $b$ will not sell any Right. This will again get him less Money and increase the price in the next Market.
\end{enumerate}
\end{proof}

Now, let us focus on coalition-proofness.
\begin{lemma}
There does not exist any coalition of traders that could increase their individual utilities by deviating from the Greedy strategy.
\end{lemma}
\begin{proof}
Let us split the proof into three parts, depending on the composition of the coalition $C$.
\begin{enumerate}
    \item $C\subset B:$ To increase the utility of buyers, they need to acquire more Good, meaning $T\setminus C$ will have less. Since $G_s=0\ \forall s\in S$ when following Greedy, $B\setminus C$ need to acquire less. The only way $C$ can accomplish that is if they don't buy Right from $B\setminus C$. But, similar to the proof of Lemma \ref{lemma: buyer deviation}, this would only lead to an increase in price, lowering the utility of $b\in C$.
    \item $C\subset S:$ The sellers have utility for Money, and since following Greedy they obtain all Money the buyers have, $S\setminus C$ needs to get less Money. The proportion in which the sellers split the Money of buyers is given by $p_s^\tau$ and $g_s$, of which they can influence $p_s^\tau$. But any price change will not decrease Money $S\setminus C$ get. Decreasing will leave some Money for the next Market, of which $S\setminus C$ gets a portion. Increasing the price will leave $C$ with extra Good, decreasing their utility.
    \item $C\subset T:$ Again, $S\cap C$ can only increase their utility if $S\setminus C$ get less Money. This can only be accomplished if $B\cap C$ don't buy from $S\setminus C$, so they accept a lower price, which is still larger than the selling price of $S\cap C$. But if the selling price is lower, and they are selling only to a subset of buyers, they cannot get more Money.
\end{enumerate}
\end{proof}

Finally, let us discuss the computational complexity of a Market utilizing the Greedy strategy.
\begin{lemma}
\label{lemma: greedy is efficient}
The Greedy strategy can be computed efficiently.
\end{lemma}
\begin{proof}
The computation of the Greedy strategy given the price of Good can be done in constant time. What remains to show is that Eq. (\ref{eq: def implicit price}) can be solved efficiently. But that is the case since the price $p^\tau\in\mathbb{R}^{+}_0$, allowing us to split the interval into at most $|B|+1$ intervals separated by points $\left\{\frac{M_b^\tau}{R_b^\tau}\right\}_{b\in B}$. In each interval, Eq. (\ref{eq: def implicit price}) can be partitioned into $|B|+1$ price intervals, where in each the computation reduces to solving a simple linear equation
\begin{equation*}
    \sum_{b\in B}M_b^\tau - (p^\tau R^\tau_b - M^\tau_b)\ \text{sign}(p^\tau R^\tau_b - M^\tau_b)= 
    p^\tau\sum_{b\in B}  R^\tau_b.
\end{equation*}
which can be solved efficiently. If its solution lies in the corresponding interval, it is a solution of Eq. (\ref{eq: def implicit price}). By Brouwer fixed-point theorem the solution is guaranteed to exist, since the price is upper bounded by the free market clearing price. The complexity is thus linear in the number of buyers.
\end{proof}
This concludes the proof of Theorem \ref{thm:greedy:equilibrial}
\end{proof}

%% file: appendix/01_proof_nonexpansive.tex
\begin{proof}
In the first part of the proof, we eliminate simple cases when the price can be equal to one. In other cases, the mapping is a contraction, which we will show in the second part of this proof.
Let us begin with a statement about uniqueness.

\begin{lemma}
For any $M^\tau_b\ge m_b, R_b^\tau \in [0,1]$, Eq. (\ref{eq: def implicit price}) has a unique solution.
\end{lemma}
\begin{proof}
For fixed Money and Right of buyers, the left-hand side is a concave, decreasing piece-wise linear function of the price. Since the right-hand side is linear, they cross at most one point. For $p^\tau=0$, the left-hand side is $\sum_{b\in B}M^\tau_b \ge \sum_{b\in B}m_b = 1$, while the right is zero, so a solution exists.
\end{proof}

It may happen that the price is one in a Market. If this occurs, it will remain unchanged for the rest of the Crisis.

\begin{lemma}
Let all traders follow the Greedy strategy. Then $p^\tau = 1 \Rightarrow p^{\tau+1} = 1$.
\end{lemma}
\begin{proof}
If $p^\tau = 1$, then
\begin{equation*}
    \sum_{b\in B}M_b^\tau - \max\{0, R_b^\tau - M^\tau_b\}= 
    1,
\end{equation*}
and in the next Market
\begin{equation*}
    \sum_{b\in B}M_b^{\tau+1} - \max\{0, p^{\tau+1} R_b^\tau - M^{\tau+1}_b\}= 
    p^{\tau+1},
\end{equation*}
but
\begin{align*}
    \sum_{b\in B}M_b^{\tau+1} 
    = 
    \sum_{b\in B}m_b + \max\{0, R_b^\tau - M^\tau_b\} 
    =1 + \sum_{b\in B}\max\{0, R_b^\tau - M^\tau_b\} = 
    \sum_{b\in B}M_b^{\tau}.
\end{align*}
However, this implies that $p^{\tau+1} = p^\tau$. For contradiction, let $B'$ be a set of poor buyers at $\tau$. The price can only be influenced by the Money poor buyers have, since $R_b^\tau$ is fixed. There are two options
\begin{enumerate}
    \item $p^{\tau+1} > p^\tau$: Then the poor buyers have more Money $\sum_{b'\in B'}M^{\tau+1}_{b'} > \sum_{b'\in B'}M^\tau_{b'}$. But that is not possible without increasing the amount of Money all buyers have, since for the rich $M^{\tau+1}_b = m_b$.
    \item $p^{\tau+1} < p^\tau$: In this case, the amount of Money the poor buyers have decreases, meaning for some $b\in B\setminus B'$, $M^{\tau+1}_b > m_b$. But for rich buyers $M^{\tau+1}_b = m_b$.
\end{enumerate}
\end{proof}

This result shows that the mapping is non-expansive if at some $\tau$, $p^\tau=1$. In other cases, we show the mapping is a contraction, i.e.
\begin{equation*}
    1 > 
    \frac{|p^{\tau+1} - 1|}{|p^\tau - 1|}. 
\end{equation*}
The rest of the proof of Proposition \ref{thm:nonmyo:main} is dedicated to proving this.
Combining Eq. (\ref{eq: def implicit price}) and (\ref{eq: next money}) we get
\begin{equation*}
    p^\tau - 1 = \sum_{b\in B}\max\{0, p^{\tau-1} R_b^\tau - M^{\tau-1}_b\} - \max\{0, p^\tau R_b^\tau - M^\tau_b\},
\end{equation*}
so the contraction condition is
\begin{figure*}
\begin{align}
\label{eq: contraction}
    1 \nonumber
    &> 
    \frac{|p^{\tau+1} - 1|}{|p^\tau - 1|} 
    =\left|
    \frac
    {\sum_{b\in B}\max\{0, p^{\tau} R_b^\tau - M^{\tau}_b\} - \max\{0, p^{\tau+1} R_b^\tau - M^{\tau+1}_b\}}
    {\sum_{b\in B}\max\{0, p^{\tau-1} R_b^\tau - M^{\tau-1}_b\} - \max\{0, p^\tau R_b^\tau - M^\tau_b\}}
    \right| \\
    \nonumber
    &=\left|
    \frac
    {\sum_{b\in B}\max\{0, p^{\tau+1} R_b^\tau - M^{\tau+1}_b\} - \max\{0, p^{\tau} R_b^\tau - M^{\tau}_b\}}
    {\sum_{b\in B}\max\{0, p^{\tau-1} R_b^\tau - M^{\tau-1}_b\} - \max\{0, p^\tau R_b^\tau - M^\tau_b\}}
    \right| \\
    &= \left| 1 - 
    \frac
    {\sum_{b\in B}\max\{0, p^{\tau-1} R_b^\tau - M^{\tau-1}_b\} - \max\{0, p^{\tau+1} R_b^\tau - M^{\tau+1}_b\}}
    {\sum_{b\in B}\max\{0, p^{\tau-1} R_b^\tau - M^{\tau-1}_b\} - \max\{0, p^\tau R_b^\tau - M^\tau_b\}}
    \right|.
\end{align}
\end{figure*}
We split the remainder of the proof into two parts
\begin{enumerate}
    \item The first option to satisfy Eq. (\ref{eq: contraction}) is
\begin{subequations}
\begin{align}
    \label{eq: app condition 1.1}
    \sum_{b\in B}\max\{0, p^{\tau+1} R_b^\tau - M^{\tau+1}_b\}&< \sum_{b\in B} \max\{0, p^{\tau-1} R_b^\tau - M^{\tau-1}_b\},\\
    \label{eq: app condition 1.2}
    \sum_{b\in B}\max\{0, p^{\tau} R_b^\tau - M^{\tau}_b\}&< 
    \sum_{b\in B} \max\{0, p^{\tau-1} R_b^\tau - M^{\tau-1}_b\}.
\end{align}
\end{subequations}
Eq. (\ref{eq: app condition 1.2}) gives
\begin{align}\nonumber
    \sum_{b\in B}\max\{0, p^{\tau} R_b^\tau - M^{\tau}_b\} &< \sum_{b\in B} \max\{0, p^{\tau-1} R_b^\tau - M^{\tau-1}_b\},\\
    \nonumber
    \sum_{b\in B}\max\{0, p^{\tau} R_b^\tau - M^{\tau}_b\} &< \sum_{b\in B} M^{\tau}_b - m_b,\\
    \nonumber
    -p^{\tau} + \sum_{b\in B}M^{\tau}_b &< -1 + \sum_{b\in B} M^{\tau}_b,\\
    \label{eq: second inequality result}
    1 &< p^{\tau}.
\end{align}
Then the Eq. (\ref{eq: app condition 1.1})
\begin{align*}
    \sum_{b\in B}\max\{0, p^{\tau+1} R_b^\tau - M^{\tau+1}_b\} 
    &< \sum_{b\in B} \max\{0, p^{\tau-1} R_b^\tau - M^{\tau-1}_b\},
    \\
    \sum_{b\in B}\max\{0, p^{\tau+1} R_b^\tau - m_b -\max\{0, p^{\tau} R_b^\tau - M^{\tau}_b\}\} 
    &< 
    \sum_{b\in B} M^{\tau}_b - m_b.
\end{align*}
Let us show that the inequality holds for all buyers, not just their sum. If $b$ is poor in $\tau$, then the inequality becomes
\begin{align*}
    \max\{m_b - M^{\tau}_b, (p^{\tau+1} - p^{\tau}) R_b^\tau\} &< 
    0,
\end{align*}
however, $M^{\tau}_b > m_b$ for a poor buyer, and using Eq.~(\ref{eq: second inequality result}) and Lemma~\ref{lemma: next price} yields $p^{\tau+1} < p^{\tau}$.

Otherwise $M^{\tau}_b = m_b$ and $p^{\tau} R_b^\tau - M^{\tau}_b \le 0$. Using Lemma~\ref{lemma: next price} we get $p^{\tau} R_b^\tau - M^{\tau}_b > p^{\tau+1} R_b^\tau - m_b$, so
\begin{align*}
    \max\{0, p^{\tau+1} R_b^\tau - m_b\} <
    0,
\end{align*}
or $b$ is rich in the next Market as well. 
\item 
Similarly, the second option is
\begin{align*}
    \sum_{b\in B}\max\{0, p^{\tau+1} R_b^\tau - M^{\tau+1}_b\} 
    &> \sum_{b\in B} \max\{0, p^{\tau-1} R_b^\tau - M^{\tau-1}_b\},
    \\
    \sum_{b\in B}\max\{0, p^{\tau} R_b^\tau - M^{\tau}_b\}
    &>
    \sum_{b\in B} \max\{0, p^{\tau-1} R_b^\tau - M^{\tau-1}_b\},
\end{align*}
where the second condition similarly reduces to $1 > p^\tau$. Focusing on the first one gives
\begin{equation}
\label{eq: first inequality, second option}
    \sum_{b\in B}\max\{0, p^{\tau+1} R_b^\tau - m_b -\max\{0, p^{\tau} R_b^\tau - M^{\tau}_b\}\} 
    > 
    \sum_{b\in B} M^{\tau}_b - m_b.
\end{equation}
Again, for a poor buyer $b$ in Market number $\tau$ we get
\begin{equation*}
    \max\{m_b - M^\tau_b, (p^{\tau+1} - p^{\tau}) R_b^\tau\} > 
    0,
\end{equation*}
which is again true. For a rich buyer we get $\max\{0, p^{\tau+1} R_b^\tau - m_b \}$, which can be zero. However, since for the poor buyers the inequality was strict, Eq.~(\ref{eq: first inequality, second option}) is satisfied.
\end{enumerate}
\end{proof}

%% file: appendix/01_proof_limit_price.tex
\begin{proof}
When investigating frustration, we can focus only on the poor buyers, for whom $R_b^\tau > M_b^\tau$. Asymptotically, Eq.~(\ref{eq: next money}) thus becomes
\begin{equation*}
M_b = \frac{m_b + R_b}{2},
\end{equation*}
corresponding to the same amount of Good they can buy, since the price is equal to one. This means their frustration is
\begin{equation*}
    f_b = \frac{R_b - \frac{m_b + R_b}{2}}{R_b} = \frac{1}{2}\left(1 - \frac{m_b}{R_b}\right) \le \frac{1}{2}.
\end{equation*}
In contrast, since in the free market $M_b = m_b$, the frustration of the poor buyers is twice as high
\begin{equation*}
    f_b = \frac{R_b - m_b}{R_b} = 1 - \frac{m_b}{R_b}.
\end{equation*}
Since the frustration of the rich buyers is zero, the overall expected frustration is asymptotically half of what it would be in the free market.
\end{proof}

%% file: sections/04_experiments.tex
\begin{figure*}[t!]
    \centering
    \includegraphics[width=0.27\textwidth]{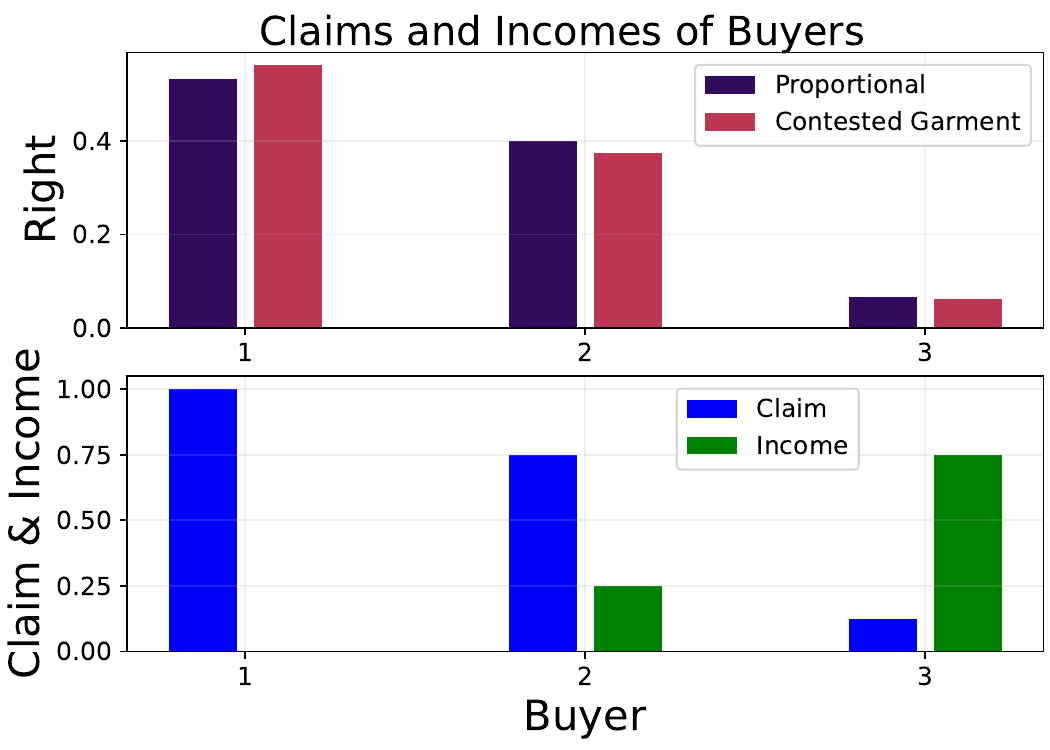}
    \includegraphics[width=0.3\textwidth]{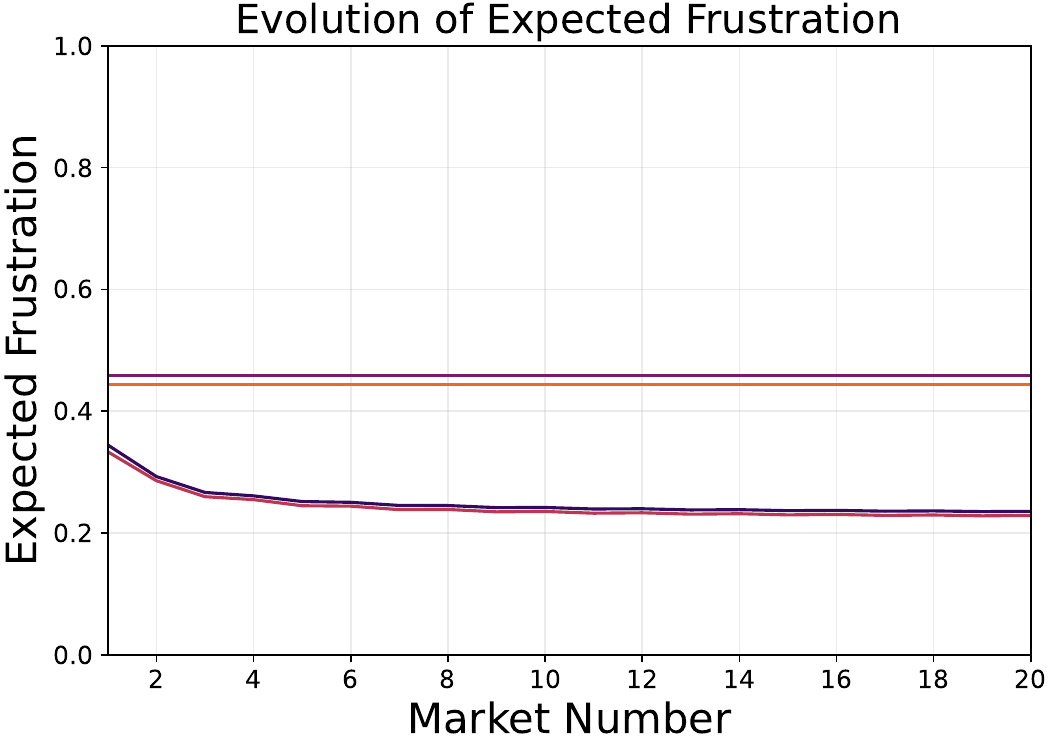}
    \includegraphics[width=0.3\textwidth]{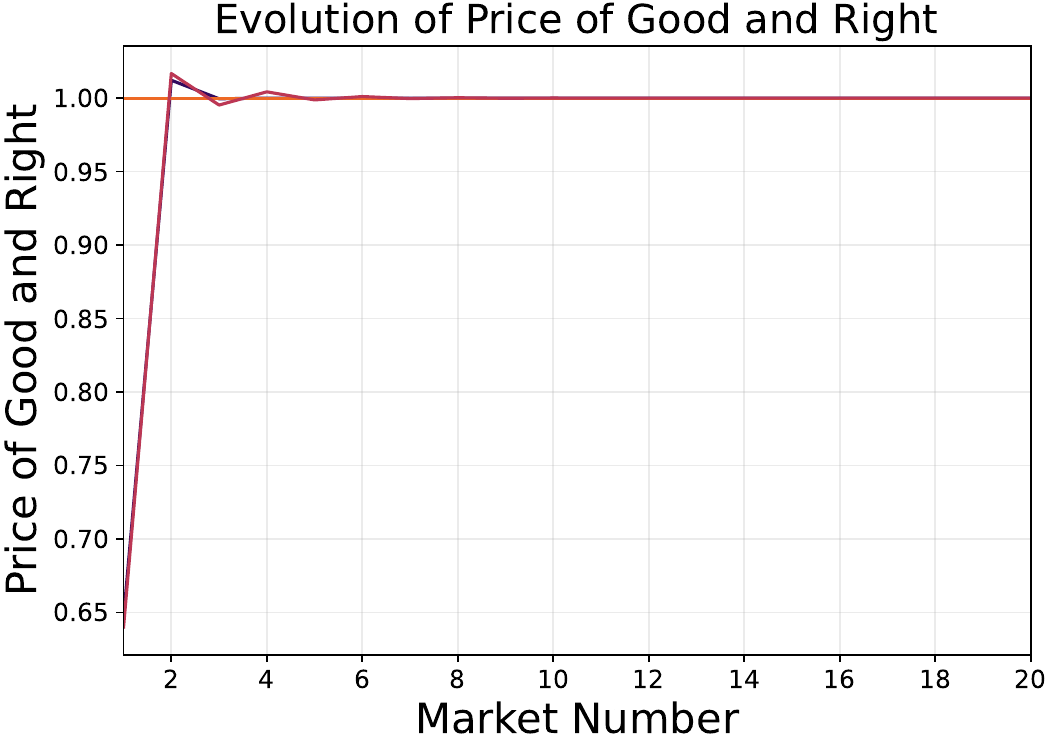}
    \\
    \includegraphics[width=0.27\textwidth]{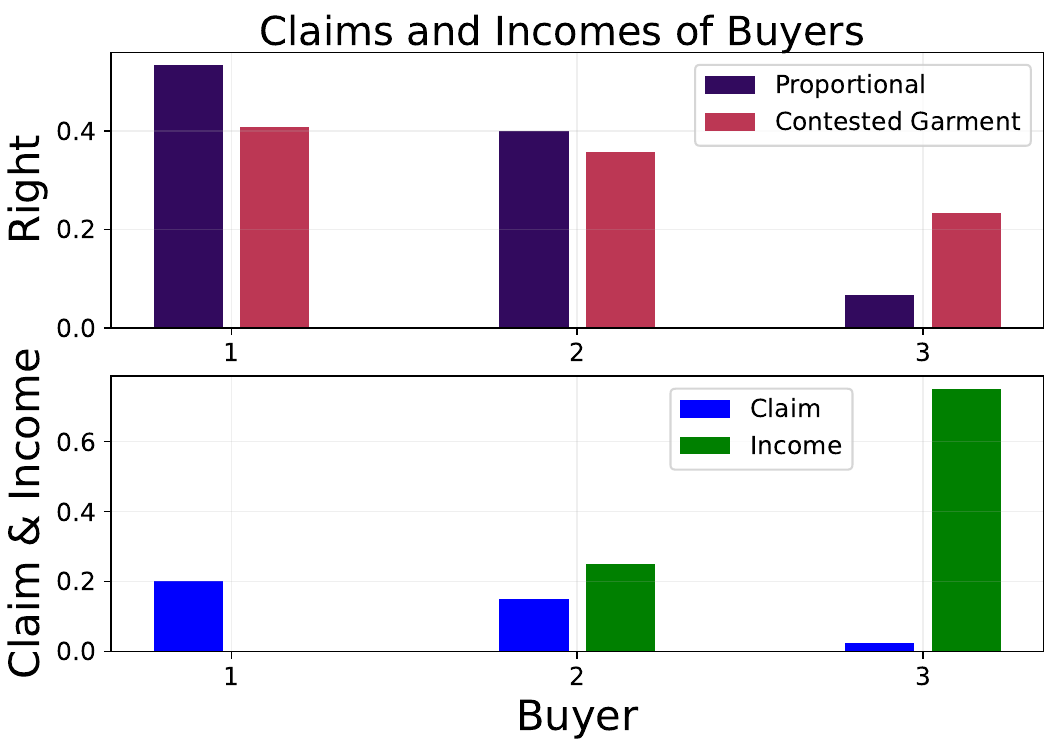}
    \includegraphics[width=0.3\textwidth]{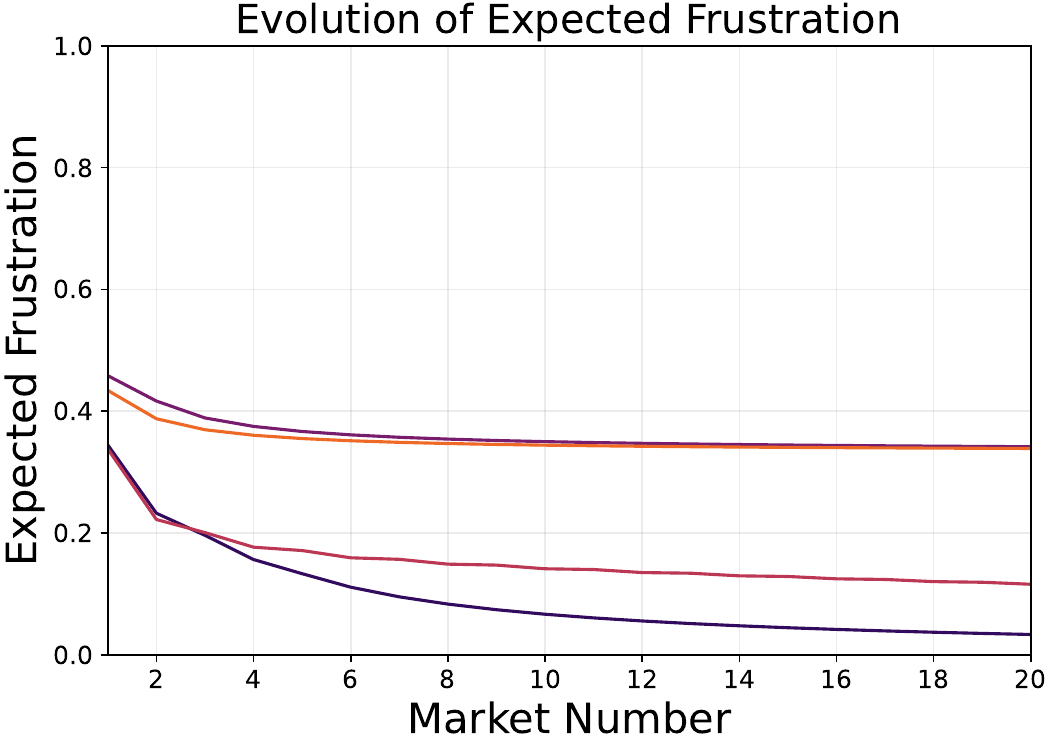}
    \includegraphics[width=0.3\textwidth]{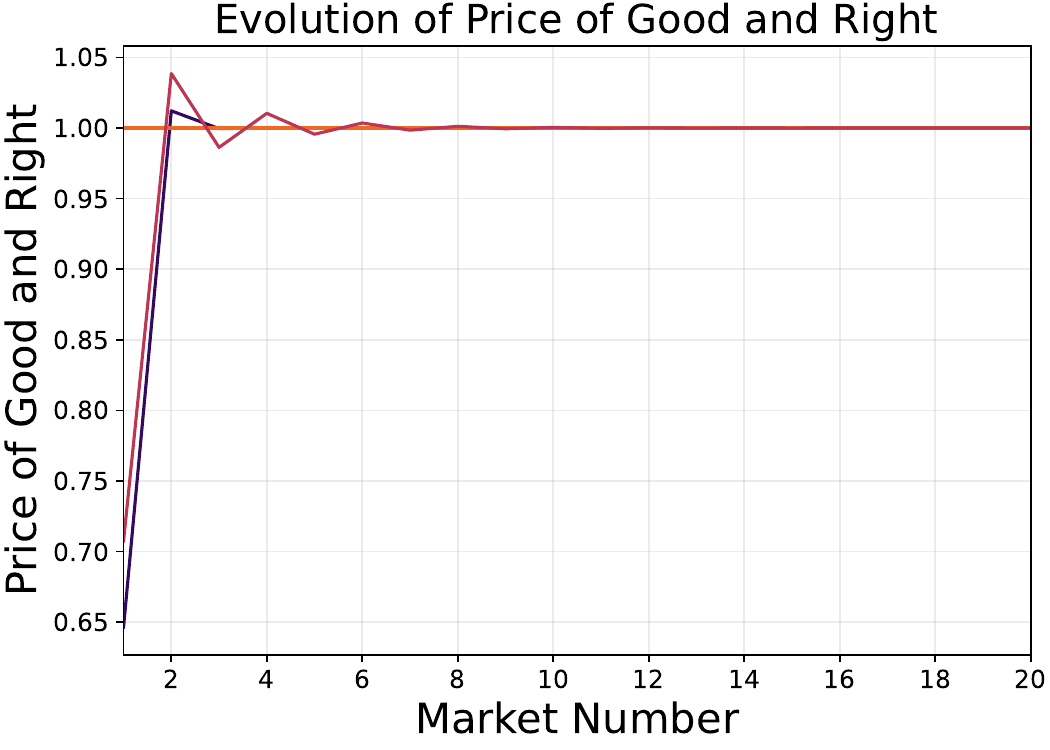}
    \\
    \includegraphics[width=\textwidth]{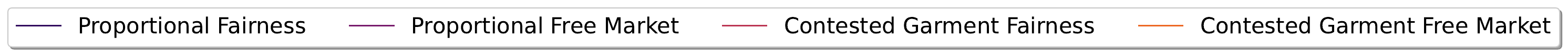}
    \caption{Comparison of the \crisisname{} with our hybrid mechanism involving trading Right (denoted ``fairness'', see Definition~\ref{def: informal market mechanism}) under the Greedy equilibrium strategy, and the free market under its equilibrium. 
    The two rows show two scenarios, which differ only by scaling the Claim five-times smaller. 
    The horizontally-left figures show the Claim $D_b$ (in blue) of each of the three buyers. 
    Colors used for the right distribution mechanisms $\phi$ (i.e. violet and red) show the corresponding amount of Right $R_b$ each buyer receives next to their Claim. 
    The horizontally-centered figures show the evolution of the expected frustration $\expectedfrustration_f^\tau$. 
    The horizontally-right figures show the evolution of the price of both Good and Right in each scenario.}
    \label{fig: experiments constant supply}
\end{figure*}

\begin{figure*}[t!]
    \centering
    \includegraphics[width=0.4\textwidth]{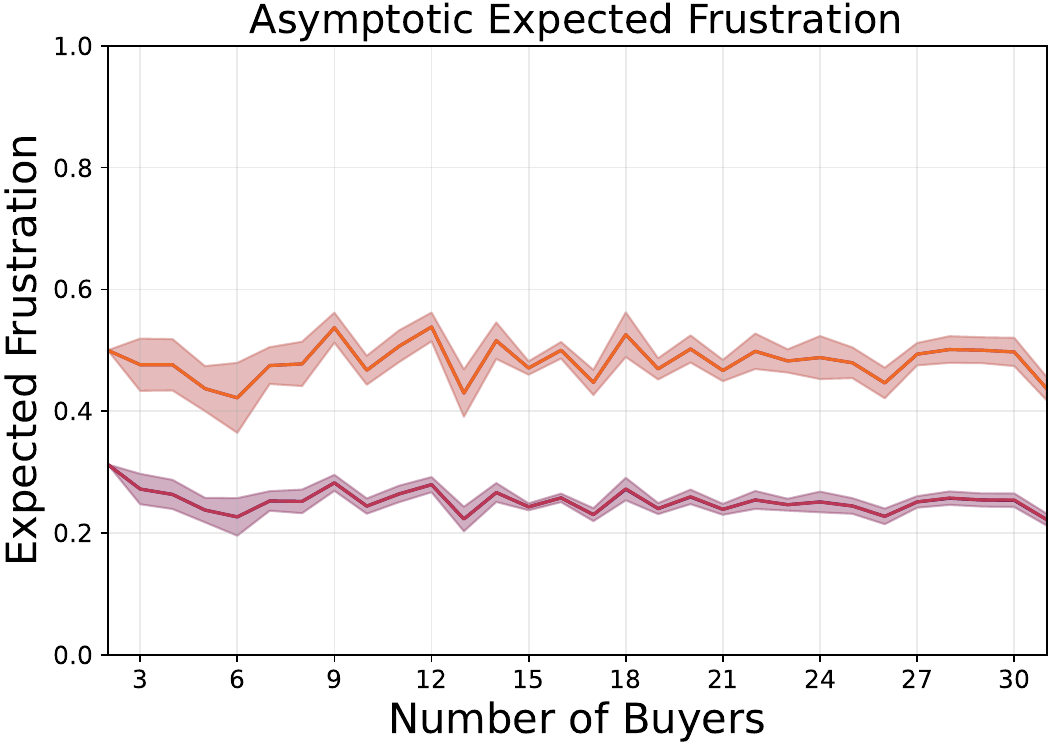}
    \includegraphics[width=0.4\textwidth]{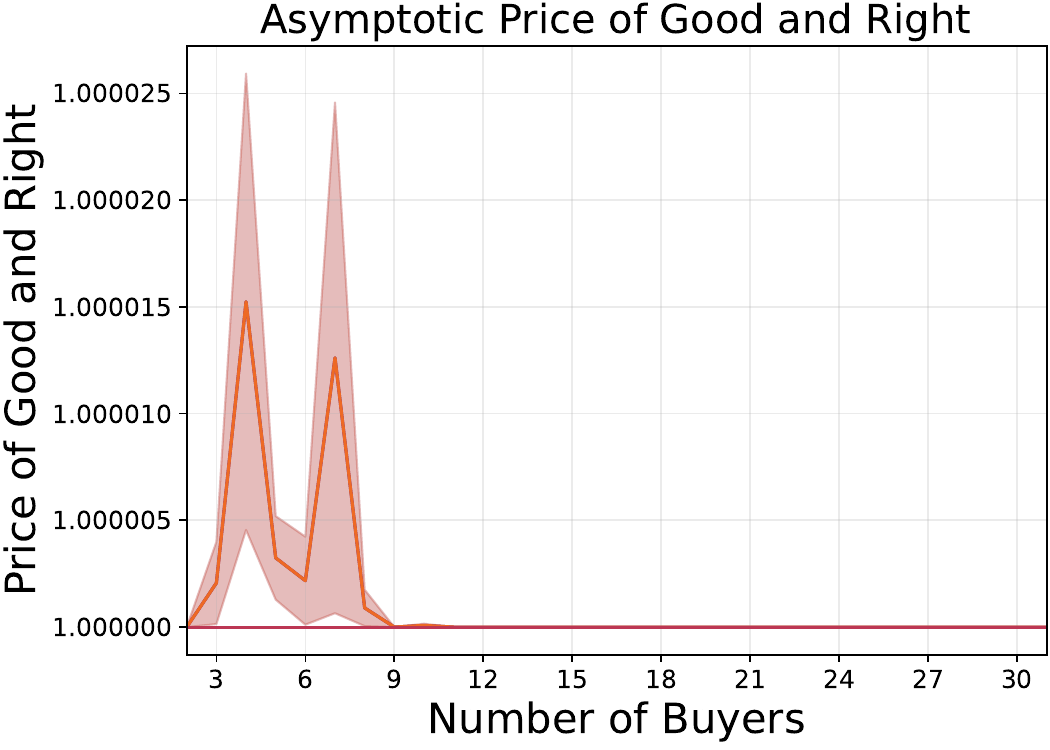}
    \\
    \includegraphics[width=0.4\textwidth]{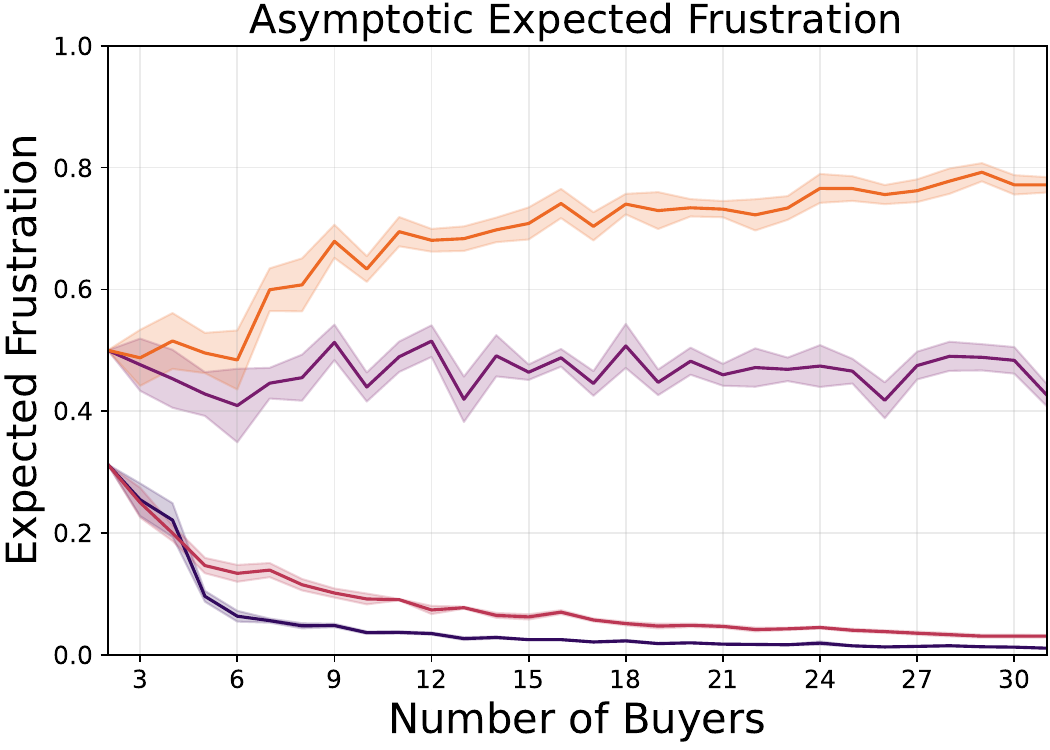}
    \includegraphics[width=0.4\textwidth]{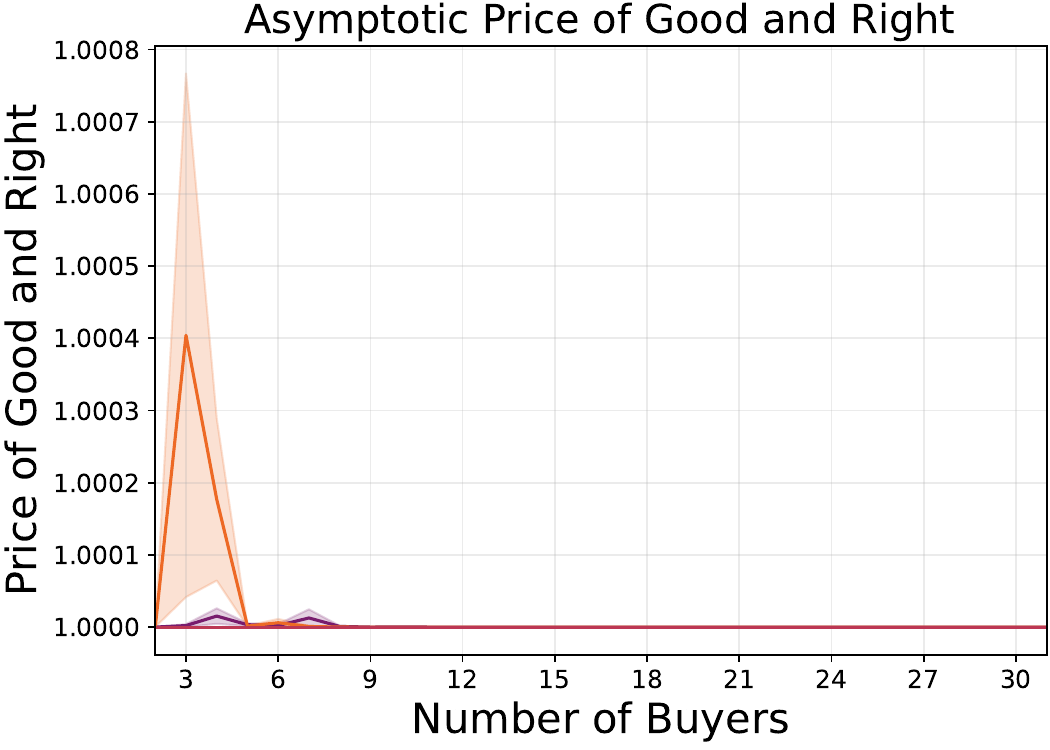}
    \\
    
    \includegraphics[width=\textwidth]{figs/legend.pdf}
    \caption{Comparison of the \crisisname{} with our hybrid mechanism involving trading Right, denoted ``fairness'', as a function of the number of buyers. To generate parameters of the \crisisname{}, we sample the Claim $D_b$ and income $m_b$ of buyers from the Dirichlet distribution. 
    To generate a scenario, we randomly order the buyers. We than set the mean Claim of a buyer inversely proportional to her position. The income is set in the same way, but with respect to the inverse ordering. Figure~\ref{fig: experiments constant supply} uses the same scheme, without applying the Dirichlet noise.
    Similar to Figure~\ref{fig: experiments constant supply}, the mean Claim is $|B|$-times smaller in the second row. The horizontally-left, resp. horizontally-right figures show an estimate of the asymptotic expected frustration, resp. price of both Good and Right. The shaded regions show standard errors. For each data-point, we sample 10 environments and perform $\mathcal{T} = 10|B|$ steps.
    }
    \label{fig: variable number of buyer}
\end{figure*}

\begin{figure*}[t!]
    \centering
    \includegraphics[width=0.3\textwidth]{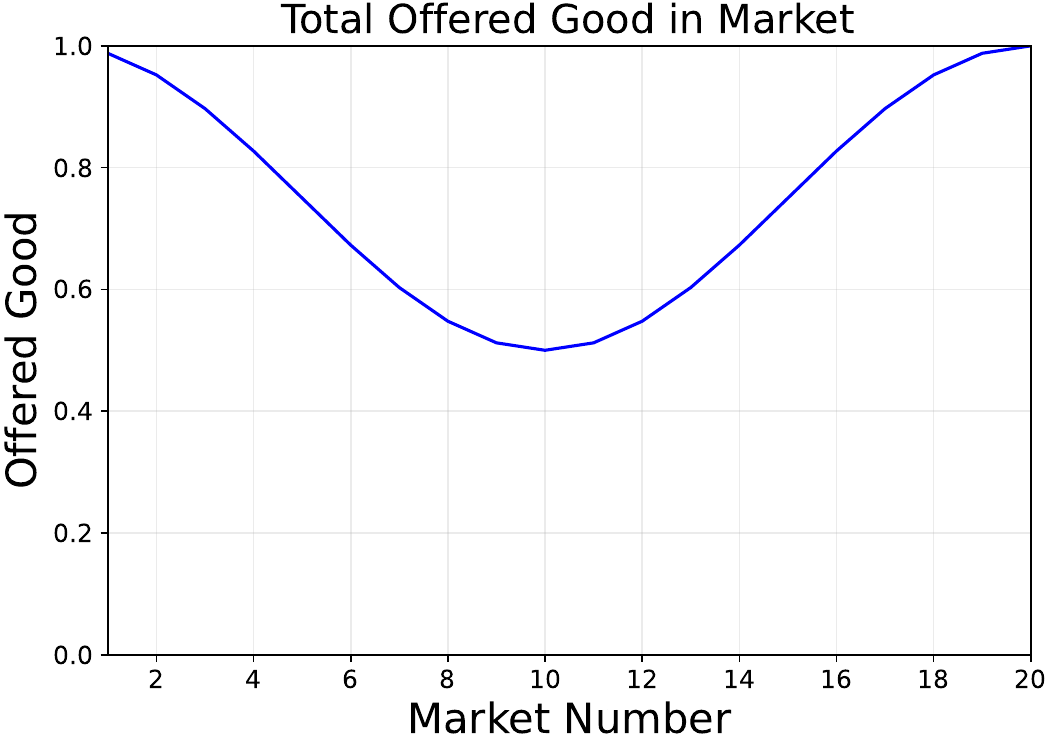}
    \includegraphics[width=0.3\textwidth]{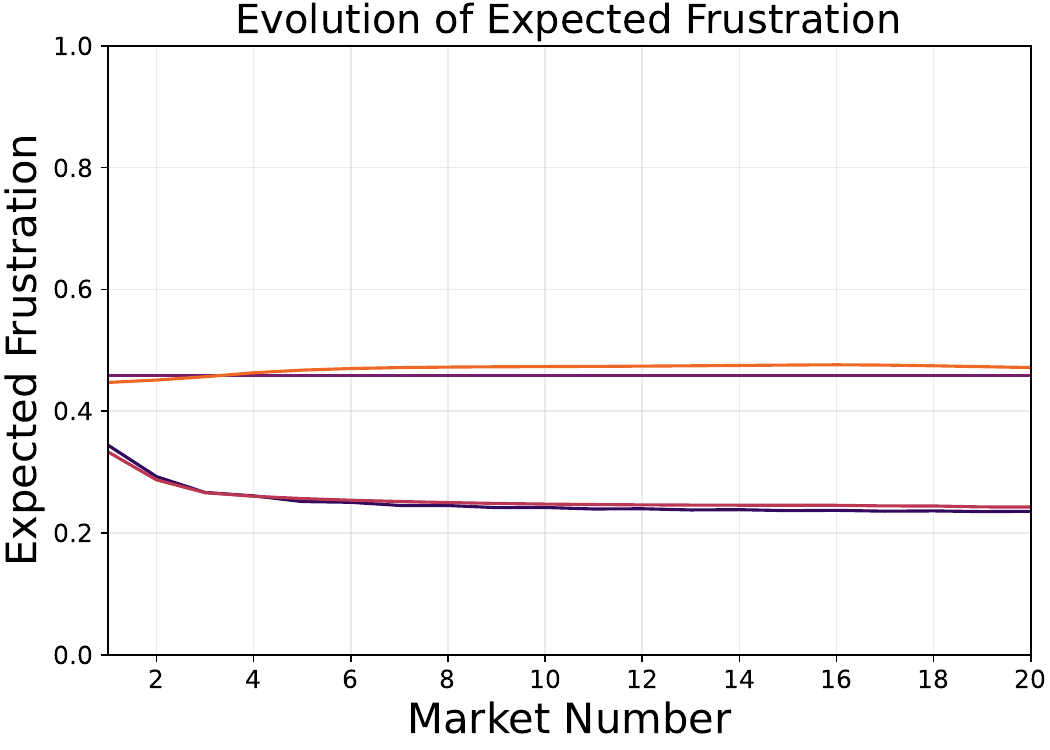}
    \includegraphics[width=0.3\textwidth]{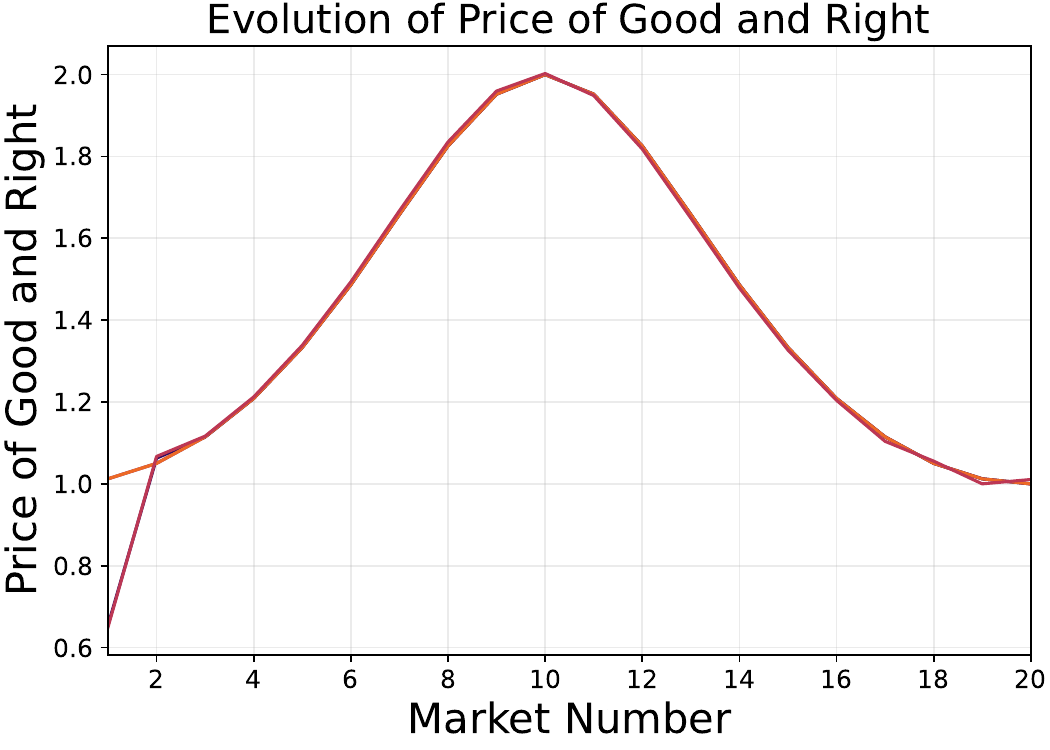}\\
    \includegraphics[width=0.3\textwidth]{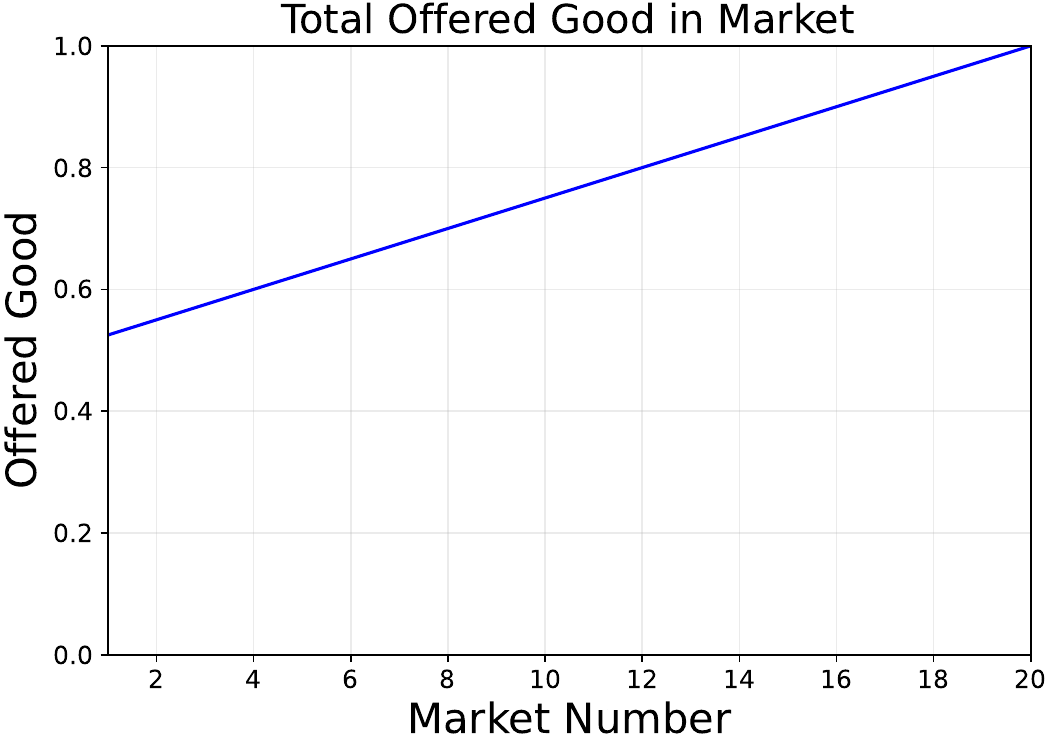}
    \includegraphics[width=0.3\textwidth]{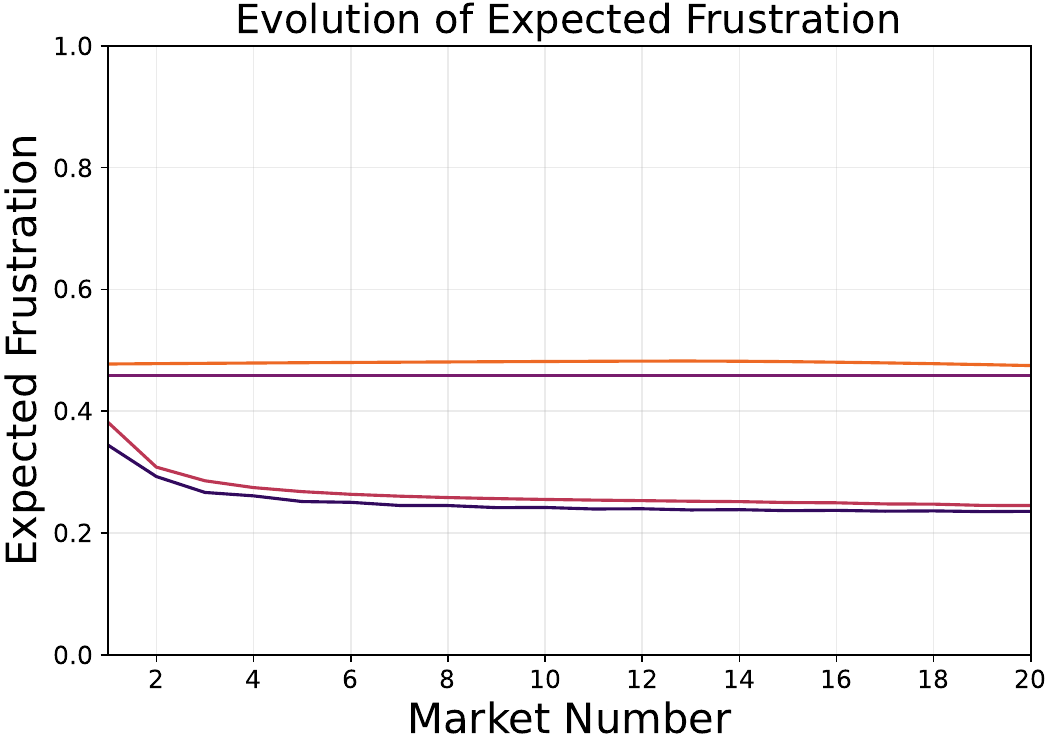}
    \includegraphics[width=0.3\textwidth]{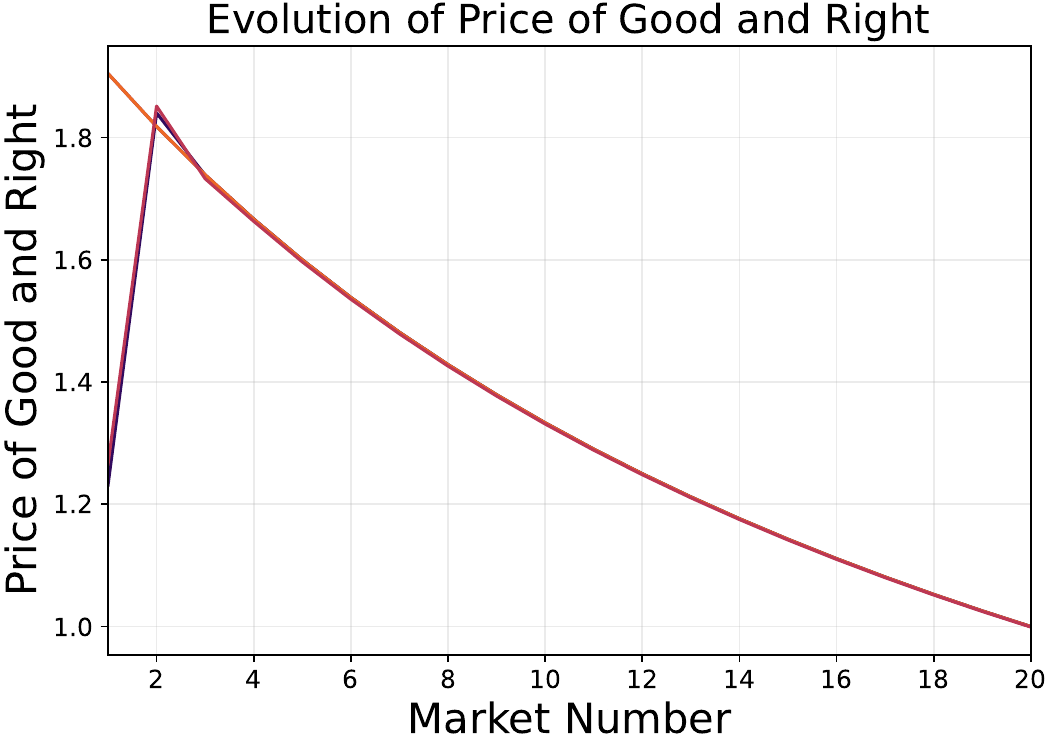}\\
    \includegraphics[width=0.3\textwidth]{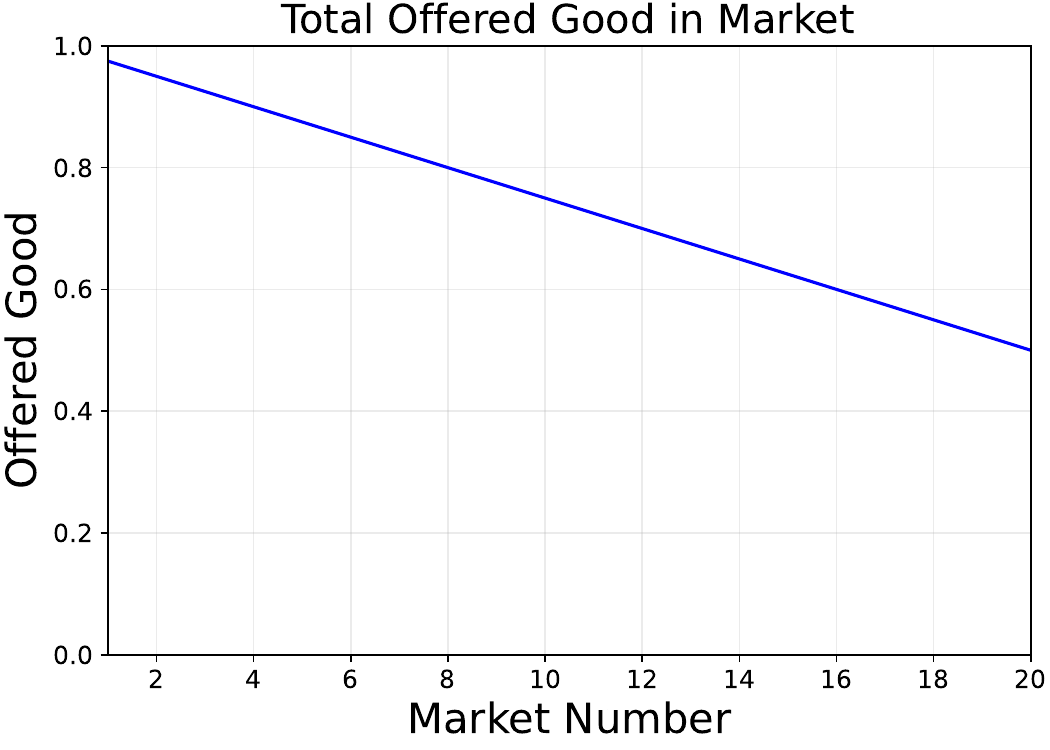}
    \includegraphics[width=0.3\textwidth]{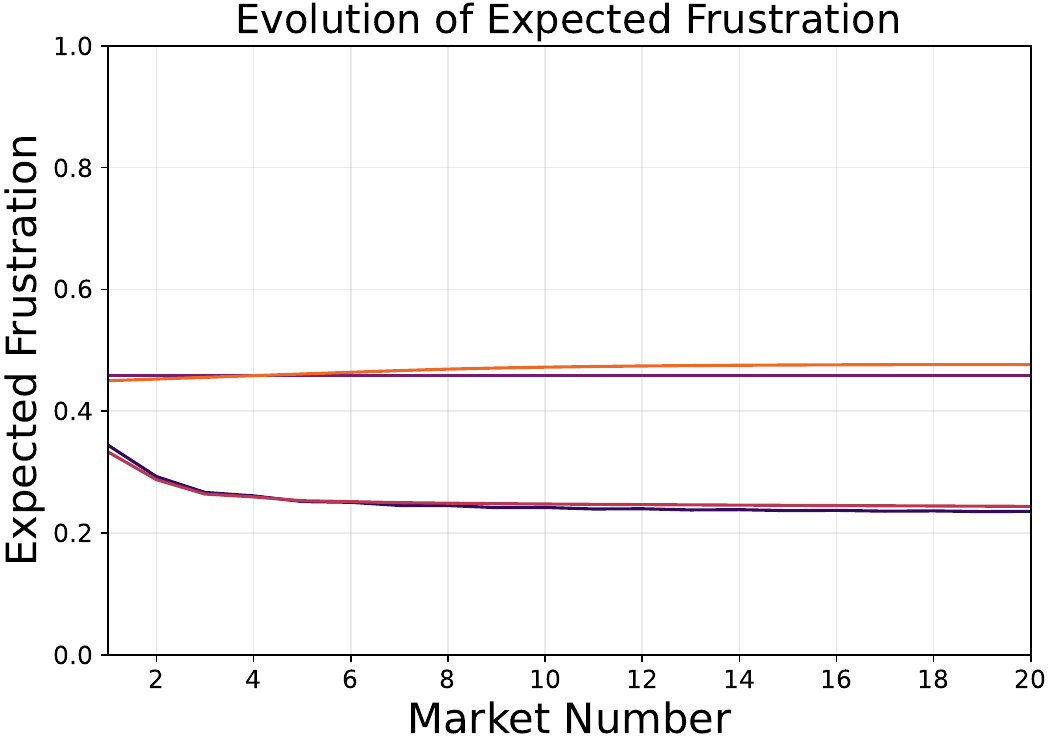}
    \includegraphics[width=0.3\textwidth]{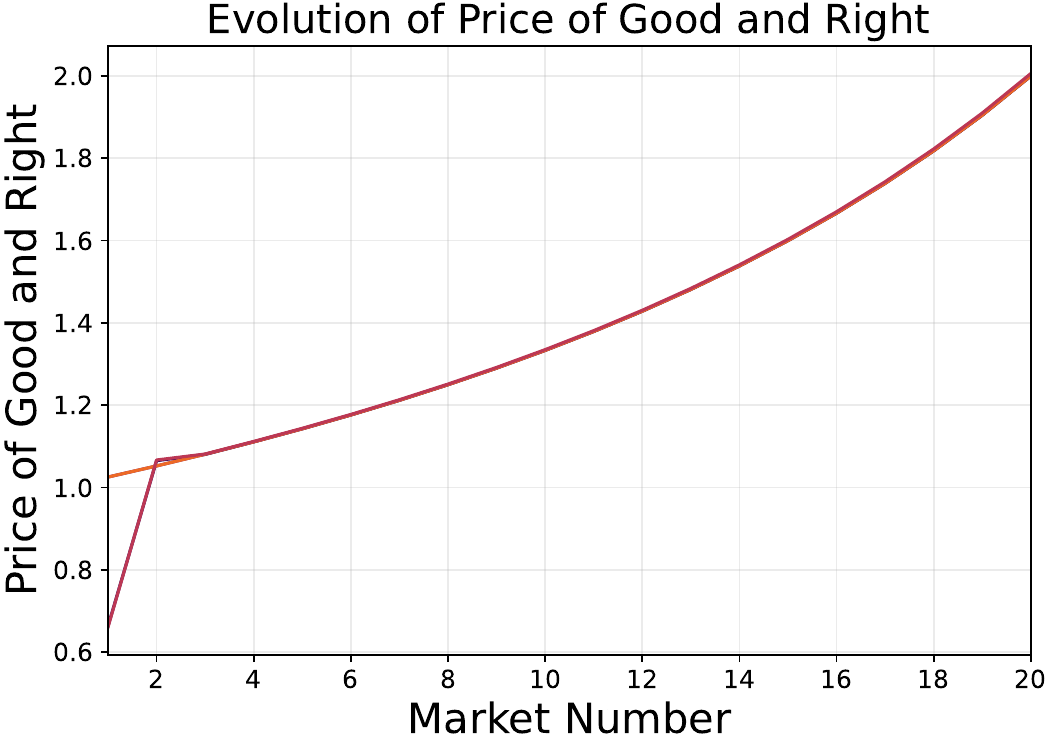}\\
    \includegraphics[width=0.3\textwidth]{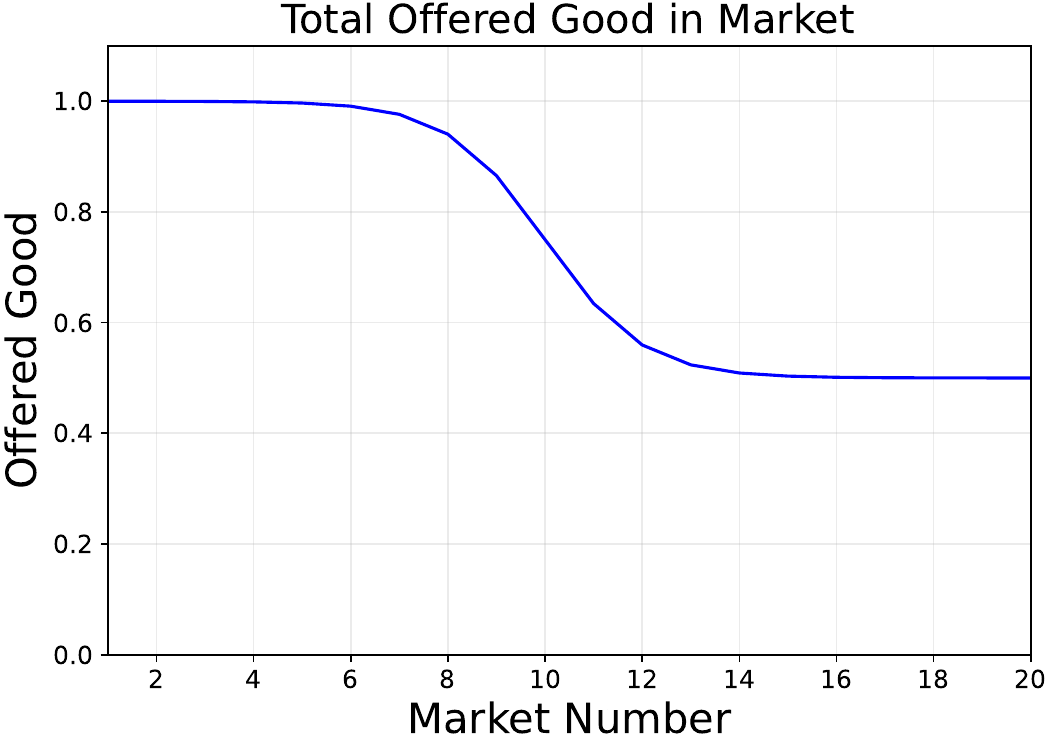}
    \includegraphics[width=0.3\textwidth]{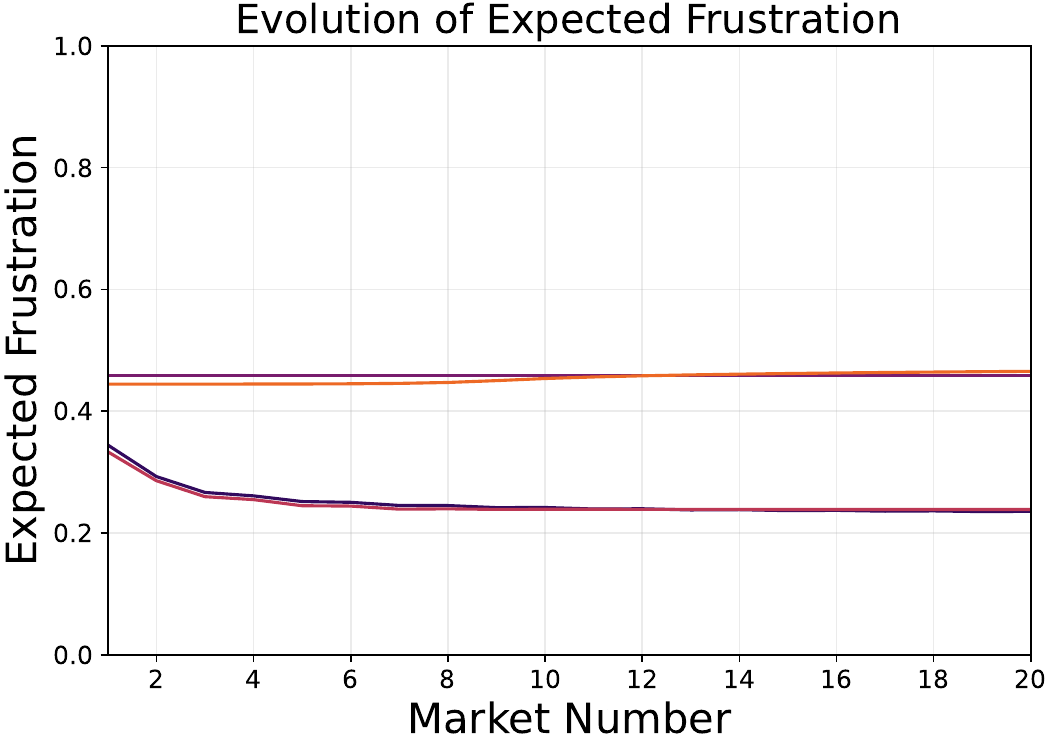}
    \includegraphics[width=0.3\textwidth]{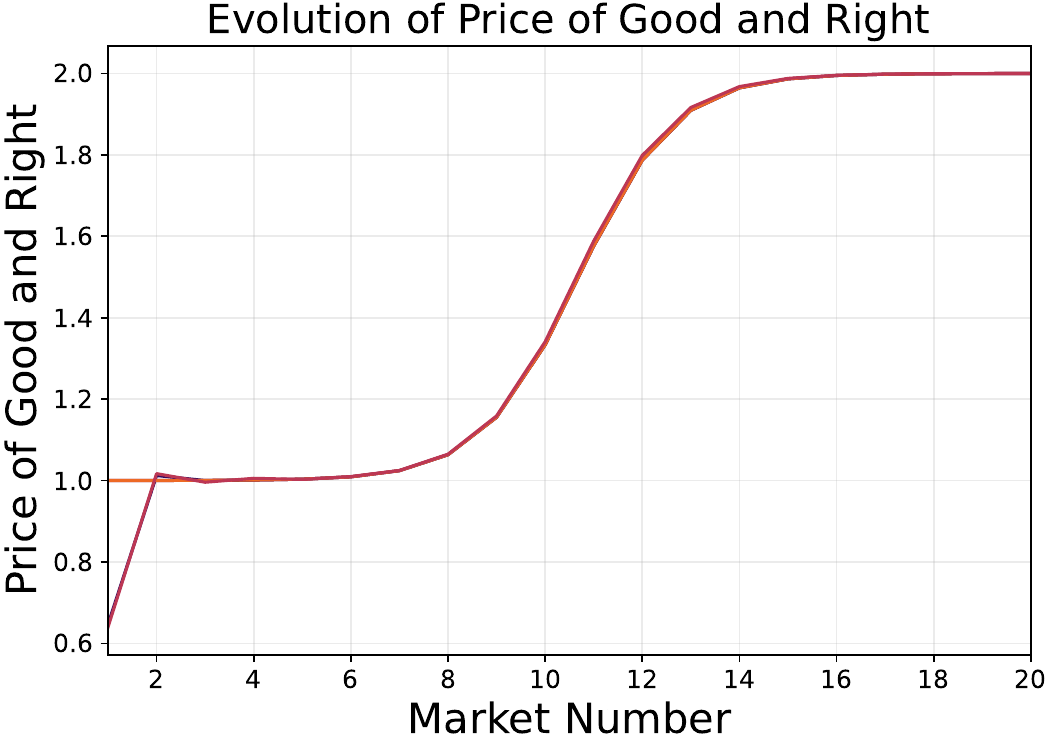}\\
    
    \includegraphics[width=\textwidth]{figs/legend.pdf}
    \caption{\crisisname~with time-variable supply of Good when our hybrid mechanism is used, compared to the free market. 
    The left column shows the total volume of Good entering the system $\sum_{s\in S}g_s^\tau$ as a function of the Market number $\tau$.
    This volume is held constant in the previous experiments.
    The centered column shows the evolution of the expected frustration $\expectedfrustration_f^\tau$ as a function of the Market number $\tau$. 
    The right column shows the evolution of the price of both Good and Right in the \crisisname. In the first row, the supply follows a cosine wave. The second and third row show linear dependence. In the fourth row, the supply follows the logistic curve.}
    \label{fig: experiments variable supply}
\end{figure*}

\begin{figure*}[t!]
    \centering
    \includegraphics[width=0.3\textwidth]{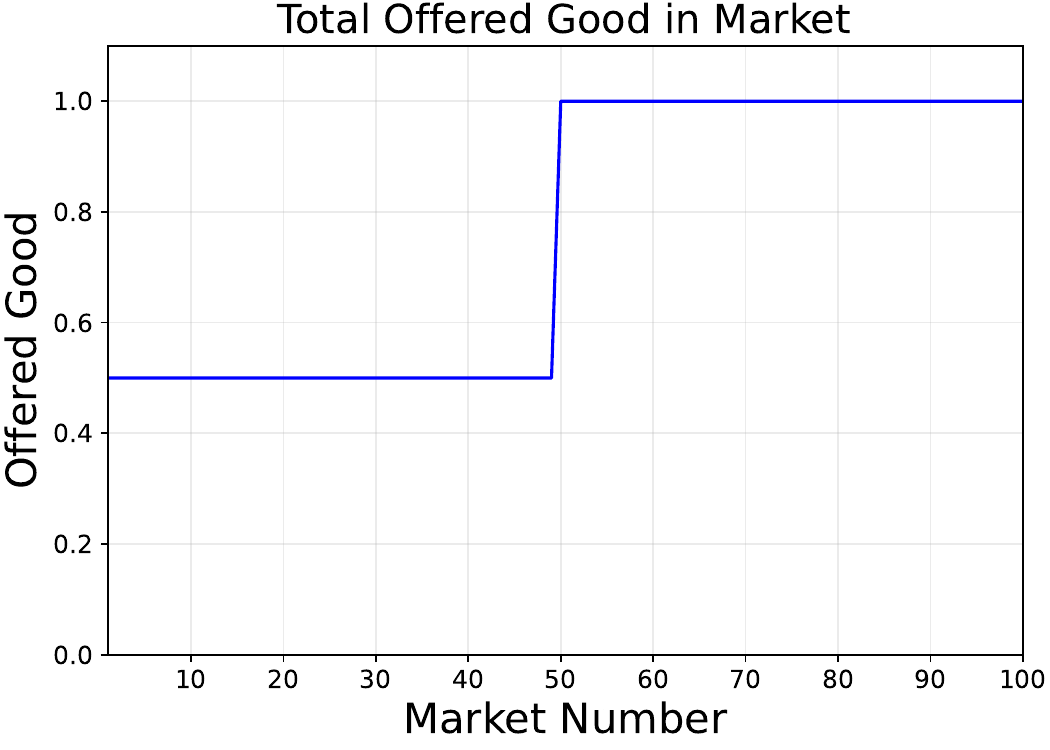}
    \includegraphics[width=0.3\textwidth]{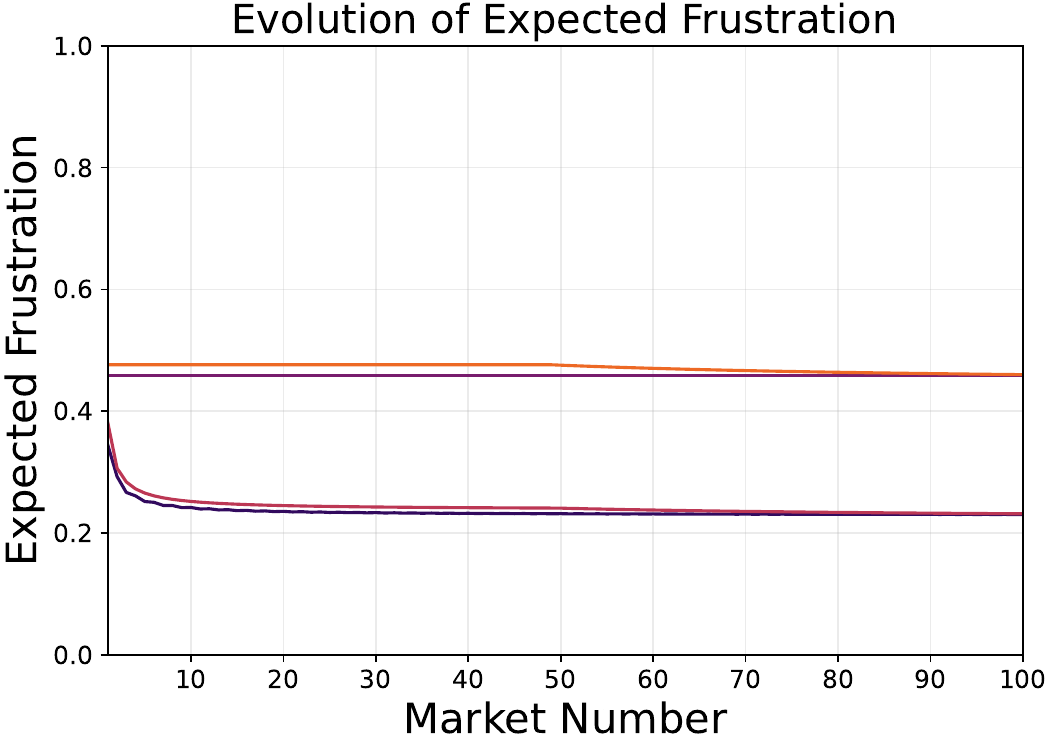}
    \includegraphics[width=0.3\textwidth]{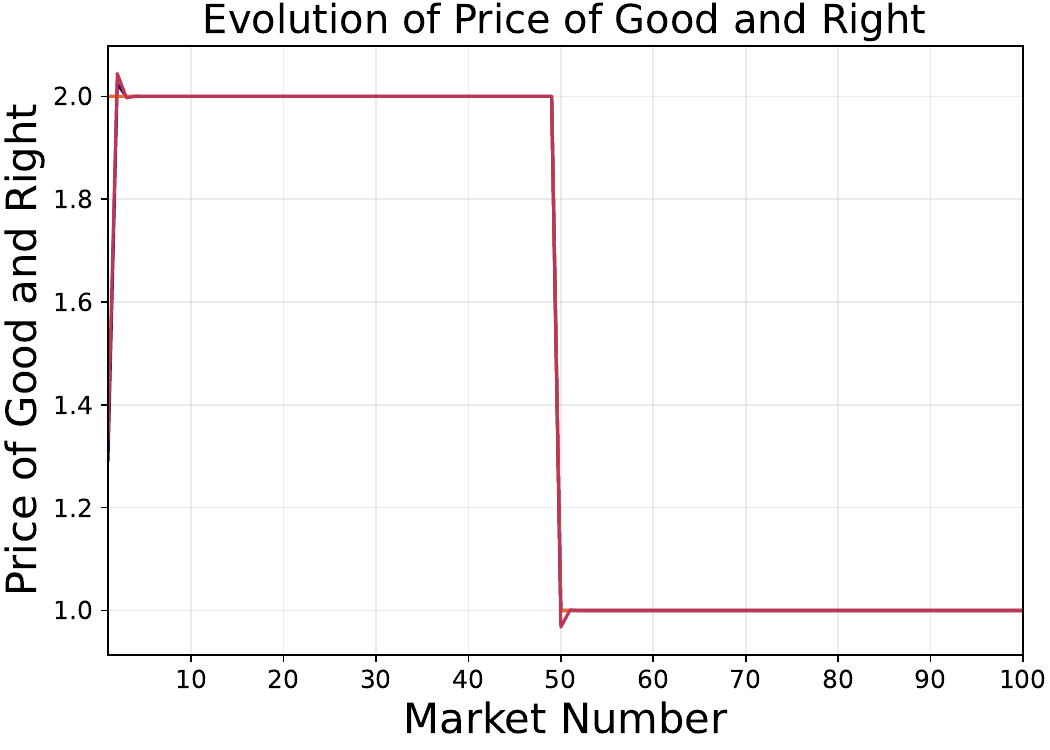}
    \\
    \includegraphics[width=0.3\textwidth]{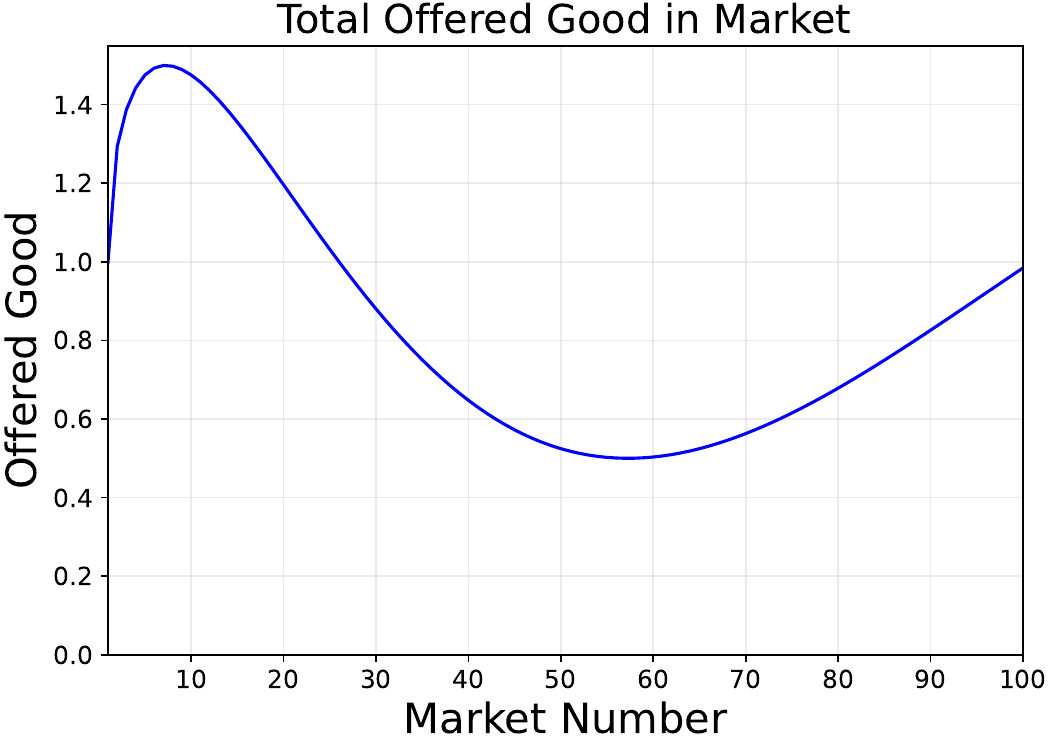}
    \includegraphics[width=0.3\textwidth]{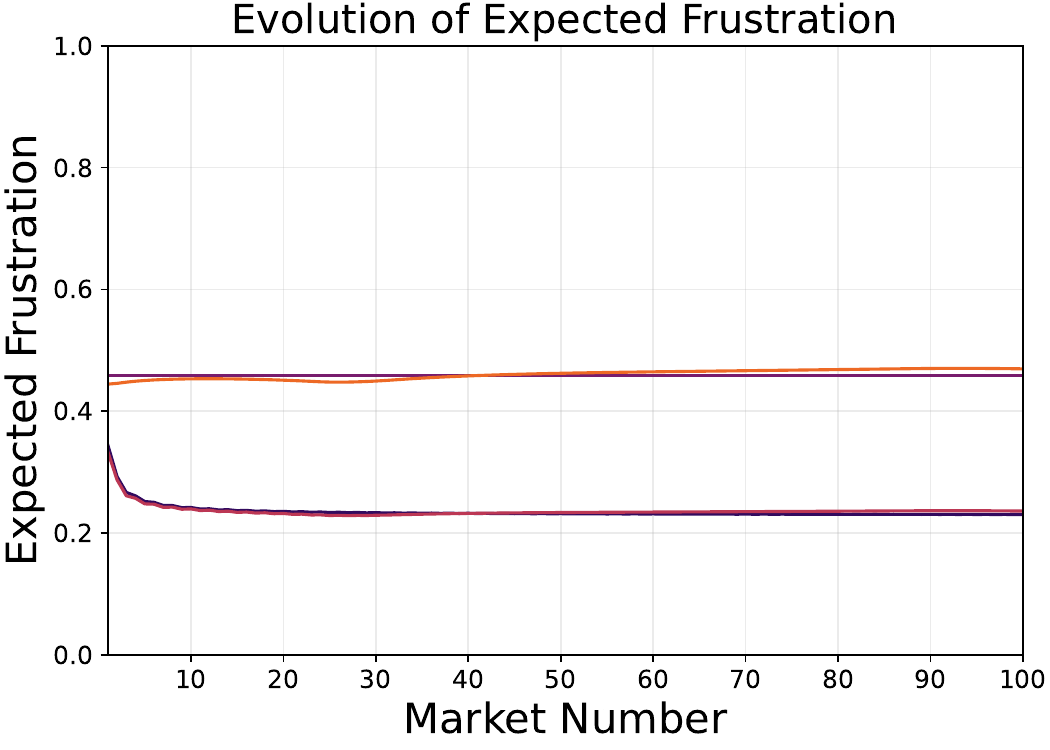}
    \includegraphics[width=0.3\textwidth]{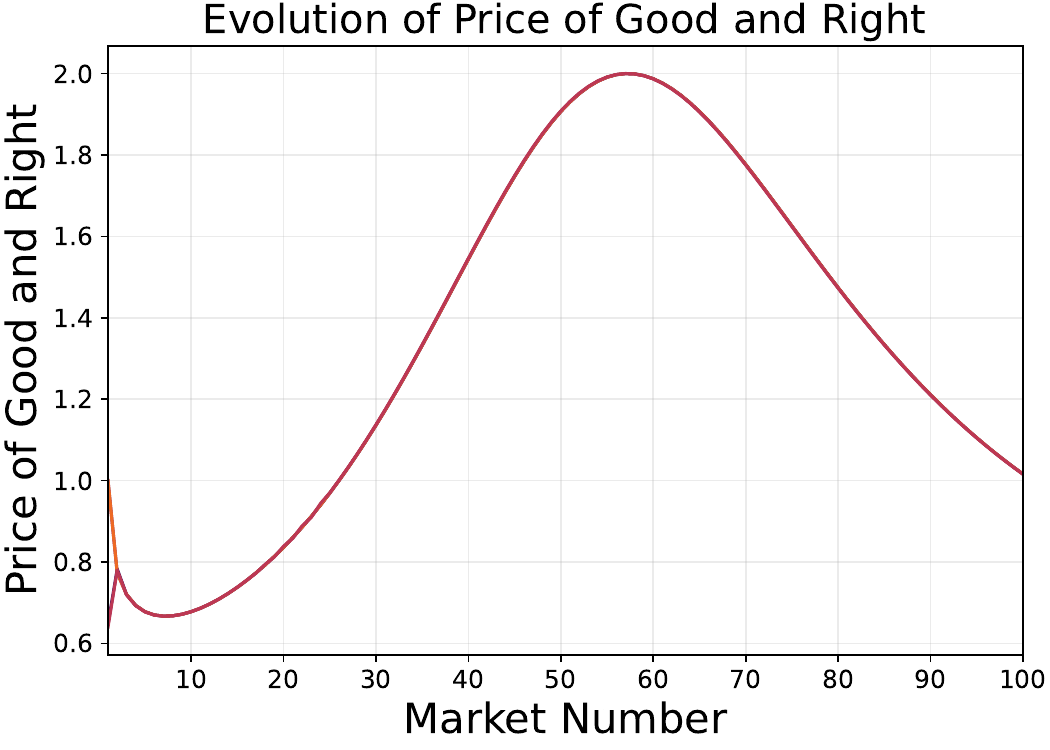}
    \\
    \includegraphics[width=0.3\textwidth]{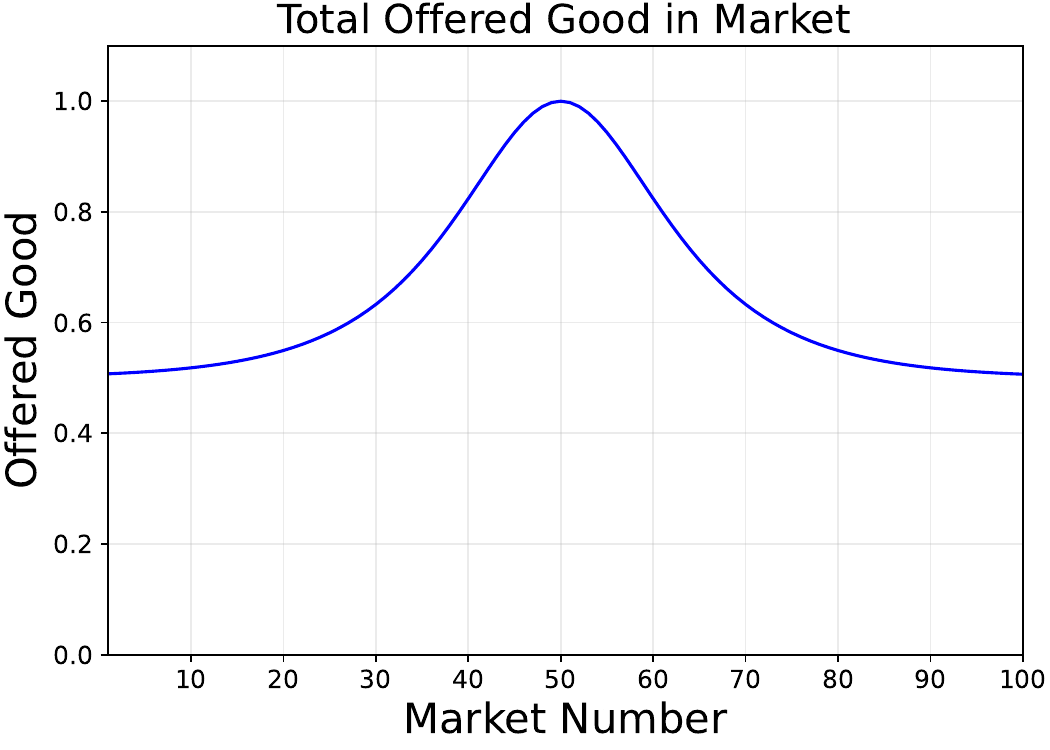}
    \includegraphics[width=0.3\textwidth]{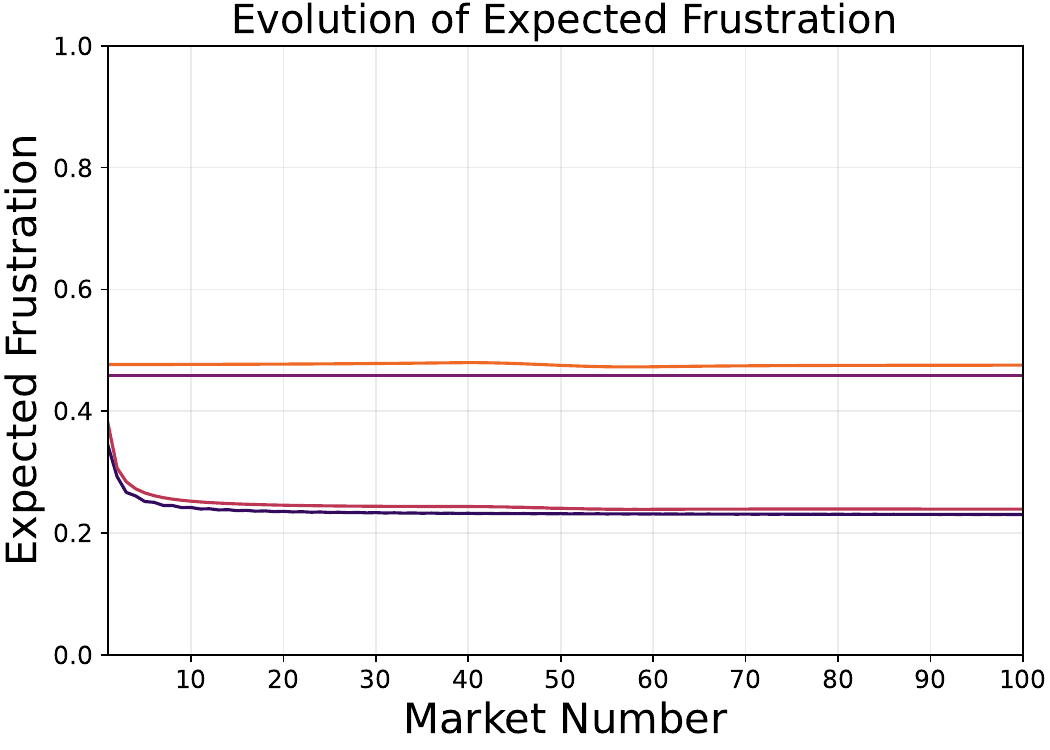}
    \includegraphics[width=0.3\textwidth]{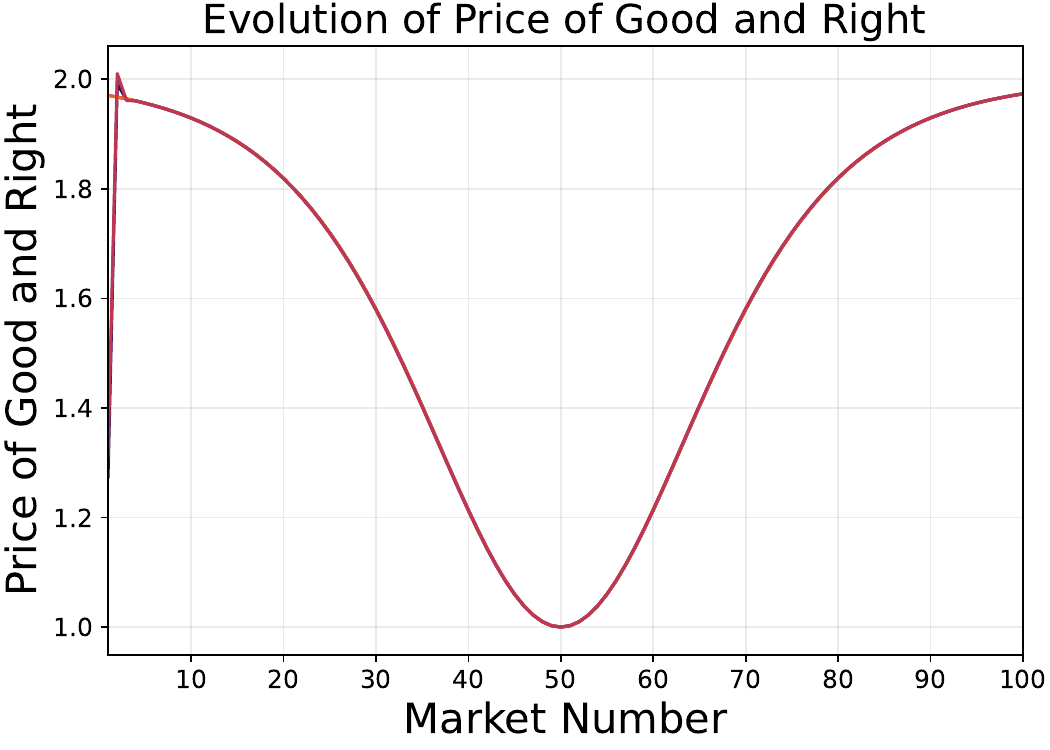}
    \\
    
    \includegraphics[width=\textwidth]{figs/legend.pdf}
    \caption{\crisisname with time-variable supply of Good when our hybrid mechanism is used, compared to the free market. 
    The left column shows the total volume of Good entering the system $\sum_{s\in S}g_s^\tau$ as a function of the Market number $\tau$.
    This volume is held constant in the previous experiments.
    The centered column shows the evolution of the expected frustration $\expectedfrustration_f^\tau$ as a function of the Market number $\tau$. 
    The right column shows the evolution of the price of both Good and Right in the \crisisname.
    The first row shows a situation where the supply of Good abruptly changes in one timestamp. The second row depict a simple model of the bullwhip effect in supply chain~\cite{lee1997bullwhip}. Finally, the third row shows the Hubbert peak supply curve~\cite{hubbert1956nuclear, calvo2017assessing}. In all cases, the price in our hybrid market closely follows the free market price. Moreover, the expected frustration is about half of the free-market value, despite the lack of theoretical guarantees of Theorem~\ref{thm:nonmyopic:poa}.}
    \label{fig: variable supply 2}
\end{figure*}

\section{Experiments}
\label{sec: experiments}

We illustrate the effect of introducing rationing via Right on a few simple scenarios.
We begin by considering situations treated analytically, where at least one of the buyers doesn't have sufficient funds to satisfy his Claim. 
Later, we study the effect of the Greedy strategies when the supply of Good changes across the \crisisname.
In this situation, our theoretical findings described in Section~\ref{sec: equilibrium} no longer hold. 

\subsection{Constant supply}\label{ssec: constant supply}

We aim to reduce the expected frustration in a situation where buyers cannot satisfy their Claim given the insufficient supply of Good in the system.
We study two such rather extreme scenarios in Figure~\ref{fig: experiments constant supply} for both the proportional distribution mechanism~\eqref{eq: proportional distribution}, and the contested garment distribution mechanism, see Appendix~\ref{app: contested garment}.
In both situations, there are two underfunded buyer who also have high Claim.

In the first scenario, the Claim is high enough to make both underfunded buyers consume all Good they buy in a Market.
Despite not receiving enough Money at the start of each Market, the underfunded buyer can sell the Right, obtaining funds for the next Market.
This manifest itself with a significantly lower frustration, which, as guaranteed by Theorem~\ref{thm:nonmyopic:poa}, decreases to half of the frustration experienced in the free market.
The price in the systems with Right starts of low as the amount of useful Money is small.
However, it quickly stabilizes near the free market clearing price of one, in agreement with Proposition~\ref{thm:nonmyo:main}.

In the second scenario, the Claim of each buyer is five-times smaller.
While there is now enough Good in the system for everyone to satisfy their Claim, the first of three buyers gets no Money at the beginning of the Market.
Thus, he is unable to buy any Good in the free market, leading to the expected frustration to approach $1/3$.
In contrast, in our hybrid system, he can sell is Right, eventually becoming non-frustrated. 
In the system with proportional fairness $f_b^\tau=0$ for $\tau \ge 4, \forall b\in B$. 
When using contested garment, since the first buyer receives less Right, this happens only for $\tau \ge 47$.

\subsubsection{Large System Limit}
To test the robustness of our system in a more realistic setting, we study the evolution of asymptotic frustration and price as a function of the number of buyers.
Our results are presented in Figure~\ref{fig: variable number of buyer}.
We again distinguish two regimes based on the ratio of total Claim $\sum_{b\in B}D_b$ and the amount of Good entering the system $\sum_{s\in S} g_s$.

In the first scenario, the total Claim is double the available Good.
In this setting, the asymptotic expected frustration stays roughly constant as we increase the number of buyers.
In agreement with Theorem~\ref{thm:nonmyopic:poa}, the asymptotic expected frustration is twice as high in the free market compared to our hybrid system.

In the second scenario, we make the Claim of each buyer $|B|$-times smaller.
This means that, for $|B| \ge 4$, there is enough Good to satisfy the Claim of all buyers.
However, the free market expected frustration when using the contested garment right distribution mechanism increases, and grows with the number of buyers.
The proportional right distribution mechanism in the free market is scaling invariant, and thus the expected frustration remains the same.
In contrast, in our hybrid system, the asymptotic expected frustration \emph{decreases} as we scale its size.
Importantly, the asymptotic price in our system remains within 1\% of the free-market value.

\subsection{Time-dependent supply}
\label{ssec: time dependent supply}

The assumption that the flow of Good and Money to the system is constant throughout the \crisisname~is critical for our theoretical analysis.
However, especially on the ``borders'' of, e. g., a distributional crisis, such assumption is not realistic.
We thus study what happens when the supply of Good changes over time, and the traders follow the Greedy strategy.
Note that Greedy is no longer an equilibrium strategy, as, for example, not selling the Good in a Market is often beneficial.

We study several modifications of the first scenario of the previous section.
We present our results in Figure~\ref{fig: experiments variable supply} and \ref{fig: variable supply 2}.
The only difference from the previous section is that the supply of each seller $g_s$ is a function of time.
In Figure~\ref{fig: variable supply 2} we study a simplified version of the bullwhip effect~\cite{lee1997bullwhip} and the Hubbert peak supply curve~\cite{hubbert1956nuclear, calvo2017assessing}, which are used to model shifts in demand and supply in markets.

Despite the lack of theoretical guarantees, the expected frustration shows a very similar trend to the constant-supply scenario.
The price of Good and Right reflects the diminishing supply, and is very similar between all systems.
This is desirable as the sellers have no incentive to avoid trading in the system with Right.

%% file: sections/05_conclusion.tex
\section{Conclusion} 


In this work, we explored the integration of buying rights into a repeated Arrow-Debreu-type market, combining elements of the (free) Arrow-Debreu market and centralized distribution. The objective was to understand how the introduction of buying rights affects the system and whether it can align the equilibrium of the free market with a more desirable, centralized solution, particularly in times of need. Buyers are assigned a new commodity -- rights -- which represents the amount of goods they are entitled to according to the centralized distribution. The system allows for the trading of both rights and goods. To assess the effectiveness of this redistribution, we used the concept of frustration, which measures the discrepancy between what a buyer actually receives and what they were entitled to. We investigate general implementations of the multi-round market and explicitly derive its coalition-proof equilibrium. 
Additionally, we show, under certain stationarity assumptions, that the expected frustration in this system is asymptotically at most $1/2$ of the expected frustration in a free market.
This demonstrates the potential benefits of integrating buying rights for improving fairness in resource allocation.
Moreover, our numerical experiments suggest that our conclusions empirically hold even in non-stationary settings.

\paragraph{Future work.}

We believe our work has three main limitations that suggest directions for future research. First, the current framework does not provide buyers with sufficient incentives to truthfully declare their claims for goods. While this may not be an issue if claims can be independently and accurately estimated, modifying the system to better incentivize truthful reporting could improve the model. Second, our theoretical results have focused primarily on scenarios with a steady, albeit small, resupply of goods and money over multiple trading rounds. Further exploration of more complex, time-dependent system dynamics would enhance our understanding of the system’s behavior under different conditions. Lastly, our results are not specific to any particular mechanism for distributing buying rights. Examining specific distribution mechanisms may influence the system dynamics and potentially improve the bounds on expected frustration. It may be possible to design tailored distribution mechanisms that meet particular fairness criteria, using approaches similar to those employed in voting mechanism design, as discussed by \citet{Koster2022}.


%% file: appendix/00_myopic_case.tex
\section{Obtaining Money for Right In the Same Market}
\label{app: myopic case}

In the main text, we have focused on the non-myopic setting where the trades aggregate their utilities over multiple timestamps.
This formulation is required because, as we defined the system above, the Markets are interdependent.
This is because the buyers are only able to access funds they obtained by selling Right in the next Market.
In this section, we briefly show that all major results also hold even in the myopic case~\cite{loebl2025market}. 
In order to make the Markets independent, we drop the constraint discussed in Remark~\ref{rem: money next market} and allow the buyers to use Money they gained for selling their Right in the current Market.

Consider the Greedy strategy defined in Section~\ref{ssec: gredy strategy}, except let the price posted by sellers be free market clearing price, or the total amount of Good in the system divided by the total amount of Money.
By construction, all Good is sold and the buyers are left with no Money. 
It also follows immediately that the frustration of each buyer is at most 1/2. 
This is because after exhausting his initial endowment of Money, he can sell half of the remaining Right at the price of Good, allowing him to buy Good with her remaining Right.

We can now check that no trader has a beneficial deviation.
The reasoning is very similar to the proof of Theorem~\ref{thm:greedy:equilibrial}.
First, the sellers can sell only a portion of their Good, keeping some for the next Market.
However, similar to Lemma~\ref{lm:nonmyo:utildec}, the price in the following Market decreases as a consequence, decreasing his utility.
Similarly, a buyer can sell less Right in a given Market. 
However, this simply leads to his Right expiring without providing Money and doesn't change the price in the following Market.
Finally, a buyer can choose to buy less Good in a Market.
This will however also not change the price in the following Market.

%% file: appendix/contested_garment.tex
\section{Contested Garment Distribution}
\label{app: contested garment}

\begin{figure*}
    \centering
    \includegraphics[width=\textwidth]{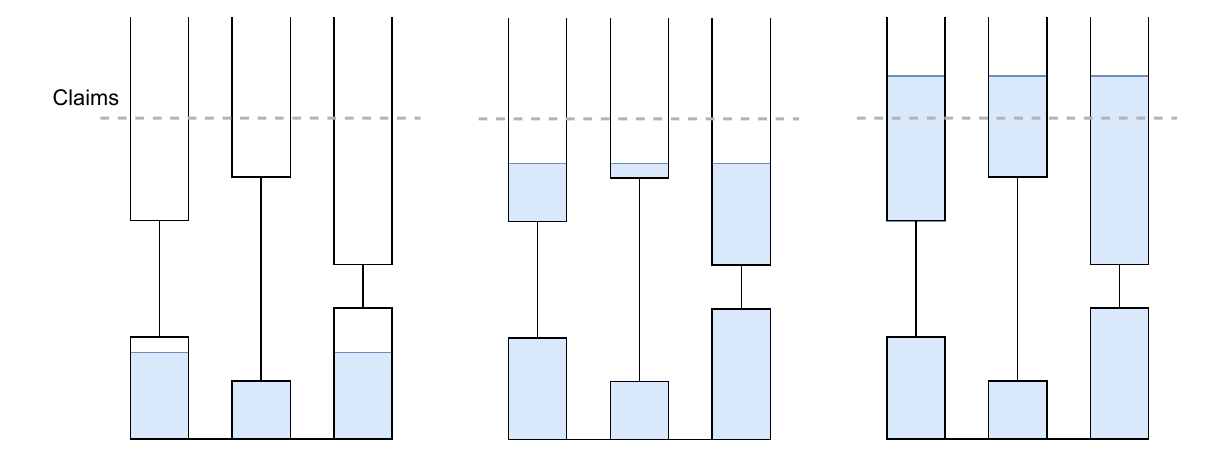}
    \caption{Illustration of the contested garment distribution. 
    The distribution can be thought of in terms of communicating vessels. 
    Each vessel is split in two part, with the volume of the bottom part equal to half of the claim of each claimant. 
    The liquid in the vessels represents the commodity being distributed.
    Therefore, the total volume of the liquid is equal to the total available commodity.
    As more liquid enters the vessels, the amount is shared evenly among the claimants, until one obtains half of his claim.
    At this point, he stop receiving more and the commodity is split among the rest equally. 
    This continues until everyone gets half of their claim.
    The situation reverses then and the distribution effectively assigns losses to the claimants in the same way as in the first phase.
    This continues until everyone's claim is satisfied.
    From that point on, the commodity is once again shared equally among all claimants.}
    \label{fig: app: contested garment}
\end{figure*}

In this section, we describe the contested garment distribution (CG)~\cite{AM}.
The goal of CG is to split, as fairly as possible, an amount of some divisible commodity among a set of claimants.
This is a non-trivial problem when the sum of the claims exceed the total available commodity.

The CG distribution can be visualized using communicating vessels, see Figure~\ref{fig: app: contested garment}.
Each vessel represents the claim of each of the claimants.
The liquid, which is poured in the communicating vessels, represents the total commodity to be distributed.
The distribution is obtained by assigning commodity equally to each claimant, whose claim is not met at least by half.
After every claimant's claim is half-met, the CG rule assigns losses to the total claim in the same way.
If everybody's claim is met, the rest of the commodity is distributed equally.

%% file: appendix/02_market_mechanism.tex
\section{Market Mechanism for Non-Myopic Traders}
\label{app: market mechanism}

In this section, we provide a formal description of the market mechanism used in Section \ref{sec:nonmyopic}. We split Definition~\ref{def: informal market mechanism} into two parts. First, we define a notion of a compatible buyer with a set of buyers and sellers as a buyer who is willing to trade with the others. Later, using this notion, we define the interaction between compatible traders.

\input{algorithms/mechanism}
\input{algorithms/market}
\input{algorithms/sequence}

\begin{cmr}{Definition (Compatible Buyer)}
Let $(v_s^\tau, p_s^\tau)$ and 

\noindent
$(w_b^\tau, q_b^\tau, \overline{v}_b^\tau, \overline{p}_b^\tau, \overline{w}_b^\tau, \overline{q}_b^\tau)$ be the bids of sellers and buyers respectively. A buyer $b\in B$ is said to be {\it compatible} with offers of $S'\subset S$ if $\overline{p}_b^\tau \ge p_{s'}^\tau$, $R_b^\tau > 0$ and $v_{s'}^\tau > 0\ \forall s' \in S'$. We denote the set of buyers compatible with offers of sellers in $S'$ as $C_1(S')$. 

Similarly, a buyer $b$ is compatible with offers of sellers in $S'$ and buyers in $B'\subset B\setminus \{b\}$ if $\overline{p}_b^\tau \ge p_{s'}^\tau$, $\overline{q}_b^\tau \ge q_{b'}^\tau$ and $v_{s'}^\tau, w_{b'}^\tau > 0\ \forall s' \in S', b' \in B'$. We denote the set of buyers compatible with offers of sellers in $S'$ and buyers in $B'$ as $C_2(S', B')$. 
\end{cmr}

\begin{cmr}{Definition (Market Mechanism)}
The market mechanism is a function $\mu: \Pi\times\mathbb{R}^{+, 2|T|+|B|}_0 \to \mathbb{R}^{+, 2|T|}_0$, written as
\begin{equation*}
    \mu(v_S^\tau, p_S^\tau, w_B^\tau, q_B^\tau, \overline{v}_B^\tau, \overline{p}_B^\tau, \overline{w}_B^\tau, \overline{q}_B^\tau, G_T^\tau, M_T^\tau, R_B^\tau) 
    = 
    (G_T^\tau, M^\tau_T).
\end{equation*}
The market mechanism we consider has two stages. In the first stage, the buyers use the Right they were assigned to buy as much Good as they desire. In the second stage, the buyers buy Good and Right in equal volume, until they buy their desired volume of either, or they have no Money left. In both stages, items offered at a lower price are traded first. When more traders offer Good or Right at the same price, they are treated as a single trader until one runs out of items for sale. 
\*

The structure of the market mechanism is outlined in the Algorithm \ref{alg: proportional market mechanism}, using the notion of compatible buyer $C_1$ and $C_2$, defined above. The overall Market can be found in Algorithm \ref{alg: market}, and the \crisisname~in Algorithm~\ref{alg: sequence}.
\end{cmr}

%% file: algorithms/mechanism.tex
\begin{algorithm*}[t]
\caption{Market mechanism}
\label{alg: proportional market mechanism}
$\Delta G_t^\tau, \Delta M_t^\tau \gets 0\ \forall t\in T$

$g \gets \text{sort}(\text{unique}(\{p_s | s\in S\}))$ \hfill \tcp{Sorted list of unique prices of Good}

$r \gets \text{sort}(\text{unique}(\{q_b | b\in B\}))$

\tcc{First stage when buyers use their Right}
\For {$p$ in $g$}{
    $S' \gets \{s |s\in S, p_s = p\}$  \hfill \tcp{Set of sellers offering Good at a given price}
    \While{$|C_1(S')| > 0$}{ 
    
    $v^\tau \gets \sum_{s\in S'} v_s^\tau$  \hfill \tcp{Total offered volume at this price}
    
    $\overline{v}^\tau \gets |C_1(S')|\ \min_{b\in C_1(S')}\{M_b^\tau / p, R_b^\tau - w_b^\tau, \overline{v}_b^\tau\}$  \hfill \tcp{Total desired and affordable volume}
    
    $V^\tau \gets \min\{v^\tau, \overline{v}^\tau\}$
    
    $\Delta G_s^\tau \gets \Delta G_s^\tau - V^\tau / |S'|,\ \forall s \in S'$
    
    $v_s^\tau \gets v_s^\tau - V^\tau / |S'|,\ \forall s \in S'$
    
    $\Delta G_b^\tau \gets \Delta G_b^\tau + V^\tau / |C_1(S')|,\ \forall s \in S'$
    
    $R_b^\tau \gets R_b^\tau - V^\tau / |C_1(S')|,\ \forall b \in C_1(S')$
    
    $M_b^\tau \gets M_b^\tau - p V^\tau / |C_1(S')|,\ \forall b \in C_1(S')$
    
    $\overline{v}_b^\tau \gets \overline{v}_b^\tau - V^\tau / |C_1(S')|,\ \forall b \in C_1(S')$
    }
}
\tcc{Second stage when buyers buy Right and Good in equal quantity}
\For {$p$ in $l$}  {
    \For {$q$ in $r$} {
        $S' \gets \{s |s\in S, p_s = p\}$  
        
        $B' \gets \{b |b\in B, q_b = q\}$  
        
        \While {$|C_2(S', B')| > 0$}{
            $v^\tau \gets \sum_{s\in S'} v_s^\tau$   \hfill \tcp{Total offered volume of Good at this price}
            
            $w^\tau \gets \sum_{b\in B'} w_b^\tau$   \hfill \tcp{Total offered volume of Right at this price}
            
            $\overline{v}^\tau \gets |C_1(S')|\ \min_{b\in C_1(S')}\{M_b^\tau / (p+q), \overline{v}_b^\tau\}$  
            
            $\overline{w}^\tau \gets |C_1(S')|\ \min_{b\in C_1(S')}\{M_b^\tau / (p+q), \overline{w}_b^\tau\}$  
            
            $V^\tau \gets \min\{v^\tau, \overline{v}^\tau, w^\tau, \overline{w}^\tau\}$  \hfill \tcp{Total volume to be traded}
            
            $\Delta G_s^\tau \gets \Delta G_s^\tau - V^\tau / |S'|,\ \forall s \in S'$
            
            $v_s^\tau \gets v_s^\tau - V^\tau / |S'|,\ \forall s \in S'$
            
            $w_b^\tau \gets w_b^\tau - V^\tau / |S'|,\ \forall b \in B'$
            
            $\Delta G_b^\tau \gets \Delta G_b^\tau + V^\tau / |C_2(S', B')|,\ \forall b \in C_2(S', B')$
            
            $M_b^\tau \gets M_b^\tau - (p+q) V^\tau / |C_2(S', B')|,\ \forall b \in C_2(S', B')$
            
            $\overline{v}_b^\tau \gets \overline{v}_b^\tau - V^\tau / |C_2(S', B')|,\ \forall b \in C_2(S', B')$
            
            $\overline{w}_b^\tau \gets \overline{w}_b^\tau - V^\tau / |C_2(S', B')|,\ \forall b \in C_2(S', B')$
        }
    }
}
\Return: $(G_T^\tau + \Delta G_T^\tau, M_T^\tau + \Delta M_T^\tau)$
\end{algorithm*}

%% file: algorithms/market.tex
\begin{algorithm*}[t]
\caption{Market}
\label{alg: market}
\SetKwInOut{Input}{input}
\Input{$G_t^\tau, M_t^\tau,\ \forall t\in T$}

$(v_s^\tau, p_s^\tau) \gets \pi_s(G_B^\tau, M_B^\tau, G_S^\tau),\ \forall s\in S$ 

$R_B^\tau \gets \phi(\sum_{s\in S} v_s^\tau, D_B)$

$(w_b^\tau, p_b^\tau) \gets \hat{\pi}_b(v_S^\tau, p_S^\tau, G_B^\tau, M_B^\tau, R_B^\tau),\ \forall b\in B$ 

$(w_b^\tau, p_b^\tau, \overline{v}_b^\tau, \overline{p}_b^\tau, \overline{w}_b^\tau, \overline{q}_b^\tau) \gets \pi_b(v_S^\tau, p_S^\tau, G_B^\tau, M_B^\tau, R_B^\tau, w_{-b}^\tau, q_{-b}^\tau),\ \forall b\in B$ 

$B_T^\tau \gets \mu(v_S^\tau, p_S^\tau, w_B^\tau, q_B^\tau, \overline{v}_B^\tau, \overline{p}_B^\tau, \overline{w}_B^\tau, \overline{q}_B^\tau, M_B^\tau, R_B^\tau)$

$u_T^\tau \gets u(G_t^\tau, M_t^\tau)$

\Return: $(G_t^\tau, M_t^\tau, u_T^\tau)$
\end{algorithm*}

%% file: algorithms/sequence.tex
\begin{algorithm}[t!]
\caption{\crisisname{}}
\label{alg: sequence}
$G_t^1, M_t^1, u_t \gets 0,\ \forall t\in T$

\For{$\tau\in \{1, \dots \mathcal{T}\}$}
{
$G_t^\tau, M_t^\tau, u_T^\tau \gets \mathbb{M}(G_T^\tau, M_T^\tau)$

$u_t \gets u_t + u_t^\tau,\ \forall t\in T$

$G_t^{\tau+1}, M_t^{\tau+1} \gets \rho(G_t^\tau, M_t^\tau),\ \forall t\in T$
}
\Return: $u_T$
\end{algorithm}

%% file: appendix/03_canonical.tex
\section{Canonical Distribution Mechanism}
\label{app: canonical solutions}
In this section, we study a solution of the \crisisname~with perhaps the simplest class of distribution mechanism. Specifically, this distribution mechanism assigns all Right to a buyer with the $n^\text{th}$ largest Claim\footnote{Ties are broken arbitrarily, but consistently.}
\begin{equation}
    \label{eq: canonical fairness}
    \varphi_{b, n}\left(\sum_{s\in S} v_s^\tau, D\right) = 
    \begin{cases}
    \sum_{s\in S} v_s^\tau & \text{if } D_b \text{ is the }n^\text{th}\text{ highest Claim},\\
    0 & \text{otherwise}.
    \end{cases}
\end{equation}
We call it the {\it n-canonical distribution mechanism}. 
Let $b_n$ be the buyer receiving all Right. 
For $\tau = 1$ we get
\begin{align}
\nonumber
    \sum_{b\in B}M_b^1 - \max\{0, p^1 R_b - M^1_b\}
    &= 
    p^1 \sum_{b\in B} R_b,\\
    \nonumber
    \sum_{b\in B}m_b - \max\{0, p^1 R_b - m_b\}
    &= 
    p^1 ,\\
    \nonumber
    1 - p^1 + m_{b_n} 
    &= 
    p^1,
    \\
    \label{eq: canonical solution}
    \Rightarrow
    \hspace{2ex}
    p^1 &= \frac{1+m_{b_n}}{2}.
\end{align}
Therefore, in the second Market, $M^2_{b_n} = m_{b_n} + \frac{1+m_{b_n}}{2} - m_{b_n} = \frac{1+m_{b_n}}{2}$ and the price is
\begin{align*}
    \sum_{b\in B}M_b^2 - \max\{0, p^2 R_b - M^2_b\}
    = 
    p^2 \sum_{b\in B} &R_b,
    \\
    1 + \frac{1+m_{b_n}}{2} - m_{b_n}
    - p^2 + m_{b_n} + 
    \frac{1+m_{b_n}}{2} - m_{b_n} 
    &= 
    p^2\\
    \Rightarrow
    p^2 = \frac{2 + 2m_{b_n}}{2} - m_{b_n} &= 1.
\end{align*}
This means $b$ will have $M^3_b = m_b + 1 - \frac{1+m_b}{2} = \frac{1+m_b}{2} = M^2_b$ and the cycle repeats. 

Any distribution mechanism $\phi$ can be constructed as a sum weighted sum of canonical mechanisms
\begin{equation}\label{eq: fairness decomposition}
    \phi(V, D) = \sum_{n=1}^{|B|} \alpha_n \varphi_{b,n}(V, D).
\end{equation}
Since Eq. (\ref{eq: def implicit price}) is quasi-linear in this composition of Right
\begin{align*}
    \sum_{b\in B}M_b^\tau - \max\{0, p^\tau (\alpha R^\tau_b + \beta \overline{R}^\tau_b) - M^\tau_b\}
    &= 
    p^\tau \sum_{b\in B} \alpha R^\tau_b + \beta \overline{R}^\tau_b,\\
    \sum_{b\in B}(\alpha + \beta)M_b^\tau - \max\{0, p^\tau (\alpha R^\tau_b + \beta \overline{R}^\tau_b) - (\alpha + \beta)M^\tau_b\}
    &= 
    p^\tau \sum_{b\in B} \alpha R^\tau_b + \beta \overline{R}^\tau_b.
\end{align*}
We want to split the maximum into two maxima, but
\begin{align*}
    \max\{0, p^\tau (\alpha R^\tau_b + \beta \overline{R}^\tau_b) - (\alpha + \beta)M^\tau_b\} 
     \le\alpha \max\{&0, p^\tau R^\tau_b - M^\tau_b\}
    \\
    + 
    &\beta \max\{0, p^\tau \overline{R}^\tau_b - M^\tau_b\},
\end{align*}
so computing each price separately gives a lower bound on the price in the composed system, which can be obtained by solving
\begin{align*}
    \alpha 
    \sum_{b\in B} M_b^\tau - \max\{0, p^\tau R_b - M^\tau_b\}
    + \beta
    &\sum_{b\in B}M_b^\tau - \max\{0, p^\tau \overline{R}_b - M^\tau_b\}\\
    &= 
    p^\tau \alpha\sum_{b\in B} R_b
    +
    p^\tau \beta\sum_{b\in B} \overline{R}_b
\end{align*}
This approach can be further generalized to an arbitrary decomposition of the distribution mechanism, leading to a lower bound
\begin{equation*}
    \sum_{n}\alpha_n\sum_{b\in B}M_b^\tau - \max\{0, p^\tau R_{b,n} - M^\tau_b\}
    =
    \sum_{n}\alpha_n p^\tau.
\end{equation*}

This construction is especially useful, since we can solve the $n$ problems separately. This can be done for the canonical distribution mechanism, leading to a lower bound on price in a \crisisname~with an arbitrary distribution mechanism.
\begin{cmr}{Proposition (lower bound on price)}
Let all traders follow the Greedy strategy. Then the price $p^\tau$ in Market $\mathbb{M}^\tau$ of a \crisisname~with distribution mechanism $\phi$ satisfies
\begin{equation*}
    p^\tau \ge 
    \frac{\sum_{b\in B}(\alpha_b + 1)M_b^\tau}{2},
    \hspace{3ex}
    \text{where}
    \hspace{3ex}
    \alpha_b = \phi_{b}(V, D).
\end{equation*}
\end{cmr}
\begin{proof}
Let $M^\tau_b$ be the amount of Money buyers have at the start of Market $\tau$. We can write the distribution $\phi$ as
\begin{equation*}
    \phi(V, D) = \sum_{n=1}^{|B|} \alpha_n \varphi_{b,n}(V, D).
\end{equation*}
where $\varphi_{b,n}(V, D)$ is the $n$-canonical distribution mechanism defined in Eq. (\ref{eq: canonical fairness}), and $\alpha_n\in [0, 1]$.

In the system with the canonical distribution mechanism we can obtain an explicit solution in a similar way to Eq.~(\ref{eq: canonical solution})
\begin{align*}
    \sum_{b\in B}M_b^\tau - \max\{0, p^\tau R_{b_n} - M^\tau_b\}
    &= 
    p^\tau, \\
    M^\tau_{b_n} - p^\tau + \sum_{b\in B}M_b^\tau  
    &= 
    p^\tau, \\
    \Rightarrow
    \hspace{2ex}
    p^\tau &= \frac{M^\tau_{b_n} + \sum_{b\in B}M_b^\tau}{2}.
\end{align*}
Weighing the prices according to the decomposition of the distribution mechanism $\phi$, we obtain the statement of the proposition.
\end{proof}